\documentclass[fleqn,usenatbib]{mnras}

\usepackage{newtxtext,newtxmath}

\usepackage[T1]{fontenc}

\DeclareRobustCommand{\VAN}[3]{#2}
\let\VANthebibliography\thebibliography
\def\thebibliography{\DeclareRobustCommand{\VAN}[3]{##3}\VANthebibliography}

\usepackage{graphicx}	% Including figure files
\usepackage{amsmath}	% Advanced maths commands

\newcommand{\OIII}{[O{\sc{iii}}]}
\newcommand{\OII}{[O{\sc{ii}}]}
\newcommand{\OI}{[O{\sc{i}}]}
\newcommand{\NII}{[N{\sc{ii}}]}
\newcommand{\SII}{[S{\sc{ii}}]}
\newcommand{\HII}{H{\sc{ii}}}

\title[Metallicities adjacent to non-SF spaxels]{The reliability of gas-phase metallicities immediately adjacent to non-star-forming spaxels in MaNGA}

\author[J. M. Scudder et al.]{
Jillian M. Scudder,$^{1}$\thanks{E-mail: jscudder@oberlin.edu (JMS)}
Aidan Khelil,$^{1, 2}$
and Jonah Z. Ordower$^{1}$
\\
$^{1}$Department of Physics \& Astronomy, Oberlin College, 110 N. Professor St. Oberlin, OH 44074, USA\\
$^{2}$Department of Physics, University of Massachusetts Amherst, 710 North Pleasant Street, Amherst, MA 01003, USA
}

\date{Accepted XXX. Received YYY; in original form ZZZ}

\pubyear{2024}

\begin{document}
\label{firstpage}
\pagerange{\pageref{firstpage}--\pageref{lastpage}}
\maketitle

\begin{abstract}
In this work, we use gas phase metallicities calculated from the Sloan Digital Sky Survey (SDSS) Mapping Nearby Galaxies at Apache Point (MaNGA) Data Release 17 (DR17) to assess the extent of potential biases in spaxels which are spatially adjacent to spaxels identified as non-star forming (non-SF) on a BPT diagram. We identify a sample of $\sim21,000$ such spaxels with calculable metallicities from the full metallicity catalogue ($\sim$1.57 million), representing a small fraction ($\sim1.3$ per cent) of the full metallicity sample.  $\sim$23 per cent of all galaxies with at least one spaxel with a calculable metallicity also contain at least one spaxel with a calculated metallicity adjacent to a non-SF spaxel, with a typical galaxy hosting 9 non-SF-adjacent spaxels. From our suite of 6 different metallicity calibrations, we find that only the metallicity calibrations based entirely on the \NII$_{6584}$/H$\alpha$ ratio are affected, showing systematic offsets to higher metallicities by up to $\sim$0.04 dex if they are located adjacent to a non-SF flagged spaxel, relative to a radially matched control sample. The inclusion of additional diagnostic diagrams (based on \OI$_{6300}$~\&/or \SII$_{6717+6731}$) is insufficient to remove the observed offset in the \NII$_{6584}$/H$\alpha$ based calibrations. Using a stricter diagnostic line on the BPT diagram removes $\sim$94 per cent of identified bordering spaxels with metallicities for all metallicity calibrations, and removes the residual offset to higher metallicity values seen in \NII$_{6584}$/H$\alpha$ calibrations. If science cases demand an exceptionally clean metallicity sample, we recommend either a stricter BPT cut, and/or a non-\NII$_{6584}$/H$\alpha$ based metallicity calibration.
\end{abstract}


%%%%%%%%%%%%%%%%%%%%%%%%%%%%%%%%%%%%%%%%%%%%%%%%%%

%%%%%%%%%%%%%%%%% BODY OF PAPER %%%%%%%%%%%%%%%%%%

\section{Introduction}

Gas-phase metallicities serve as important tracers of the chemical and dynamical history of galaxies, marking the enrichment of the interstellar medium (ISM) through the cumulative impact of generations of stars forming and subsequently explosively re-introducing metals into the ISM.  
While early work relied on targeted spectral observations of galaxies, statistically driven studies have become much more feasible with the advent of large statistical spectrographic surveys, such as the Sloan Digital Sky Survey \citep[SDSS;][]{York2000}. In more recent years, large integral field spectrographic (IFS) surveys have been completed, such as the Calar Alto Legacy Integral Field Area Survey \citep[CALIFA;][]{Garcia-Benito2014, Sanchez2012}, the Sydney-AAO Multi-object Integral field spectrograph (SAMI) Galaxy Survey \citep{Bryant2015, Allen2015, Croom2021}, and the SDSS Mapping Nearby Galaxies at Apache Point survey \citep[MaNGA;][]{Bundy2015}. These IFS surveys have increased both the data volume and complexity. While the total number of individual galaxies in these surveys is lower than in previous single-fibre surveys, the radial coverage provided by the spatial pixel (spaxel) approach of an IFS survey results in a substantial net increase in the total number of spectra, while also providing spectra at larger radii than in the single-fibre generation, which largely placed spectra in the nuclear (and brightest) region of the galaxy \citep{York2000, DR7}.

As the data available have evolved, our understanding of gas-phase metallicities has done the same. An early correlation between the nuclear gas phase metallicity and the stellar mass of the galaxy \citep[MZR; e.g.,][]{Lequeux1979}, confirmed with statistical samples \citep{Tremonti2004, Sanchez2017, Zahid2017}, remains when information across the disk of the galaxy is considered, now visible as a resolved stellar mass surface density ($\Sigma_*$) -- metallicity correlation \citep{Rosales-Ortega2012, Sanchez2013, Barrera-Ballesteros2016, Baker2023a}. Studies have also found that the MZR changes its vertical scaling with cosmic time, with higher redshift galaxies found with typically lower metallicities \citep[e.g.,][]{Erb2006, Maiolino2008, Mannucci2010, Zahid2011, Zahid2014a, Maiolino2019}, bolstering the understanding that generations of stellar evolution impact the metal content visible in the ISM of a galaxy. 
The typical spiral galaxy also tends to have a negative metallicity gradient in both targeted \citep[e.g.,][]{Rich2012, Sanchez-Menguiano2016, Belfiore2017}, and IFS surveys \citep{Sanchez2014, Perez-Montero2016, Belfiore2017}, lending support to an inside-out formation model for galaxies, where the central regions of the galaxy are older than the outskirts, and correspondingly have had more time to enrich the regions of hydrogen gas closer to the centre of the galaxy. 

Departures from this negative metallicity gradient have been used to infer the presence of some kind of redistribution of gas, either through a galactic fountain depositing enriched gaseous material in the outer regions of the galaxy \citep{Belfiore2014,Belfiore2017}, or by through low metallicity gas falling inwards. This latter could be in response to secular processes such as bars \citep{Jogee2005, Chen2023} or to a large scale gravitational perturbation of the galaxy, such as during an interaction, where both observations \citep[e.g.,][]{Rich2012, Scudder2012b, Thorp2019} and simulations \citep[e.g.,][]{Torrey2012, Scudder2012b, Bustamente2018} agree nuclear metal dilution is present at the $\sim 0.05$ dex level. 

 Metallicity calibrations have as a fundamental theoretical underpinning that the strong emission lines used as input are generated by photoionization regions surrounding young stars \citep[for a review, see][]{Kewley2019}, which are hot, UV-luminous, and temporary sources of illumination. This assumption means that for the creation of any metallicity sample, part of the initial quality control is to eliminate any regions of a galaxy (or, in the case of single-fibre data, any \textit{galaxies}) which are dominated by other sources of radiation. For metallicity work in particular, the shift towards IFS surveys has the advantage of permitting metallicity calculations to be undertaken for the outer regions of a galaxy, even when the central region is identified as dominated by some non-photoionization source of radiation.

A frequent culprit of this assumption-breaking radiation is the presence of an Active Galactic Nucleus (AGN); the radiation fields produced by an AGN are harder, resulting in far more high energy photons being released into the ISM than a UV bright star would be capable of. In turn, these high energy photons are able to penetrate more deeply into neutral hydrogen clouds, increasing the thickness of a partially ionized zone of gas surrounding the radiation source. For a photoionized region produced by an O or B type star, this partial ionization zone is typically thin \citep{Stromgren1939}, and to eliminate these regions with proportionally more voluminous partially ionized zones, a diagnostic diagram plotting easily detected emission line ratios against each other is typically used, such as that of \citet[BPT]{Baldwin1981}, which plots \OIII$_{5007}$/H$\beta$ against \NII$_{6584}$/H$\alpha$. \OIII$_{5007}$~is produced in regions of fully ionized gas, which can be powered either by UV radiation from young stars, or by an AGN. However, \NII$_{6584}$~is preferentially produced in partially ionized regions, and so will be brighter when the partially ionized regions are larger due to the release of high energy photons by (for example) an AGN. 
The results of unaddressed AGN contamination will generally inflate the metallicity estimated, by inflating the flux of emission lines produced within those partially ionized regions \citep[e.g.,][]{Kewley2019}.

While the parameter space of the BPT diagram was delineated relatively early, the exact placement of the division between the ``star-forming'' wing and the ``AGN'' wing of the diagram has varied from study to study and science case to science case. For instance, \citet{Kewley2001} offered a curve which was meant to indicate the highest possible location a stellar source could be placed, and is often used if a particularly clean AGN sample is desired. On the other extreme, \citet{Stasinska2006} is a much more conservative cut, designed to produce a clean star forming sample, with minimal AGN or other non-stellar contamination. Splitting the difference between these two diagnostics is that of \citet{Kauffmann2003}, which is an empirical curve designed to cut through the middle of the SDSS Data Release 1 \citep{Abazajian2003}, which contained over 55,000 spectra classifiable on the BPT diagram. 

Other works have assessed whether the BPT diagram is able to continue to distinguish hard ionization sources from lower energy sources in resolved data; \citet{Law2021} undertook this using the MaNGA MPL-11 data, and found that the diagnostic lines of \citet{Kauffmann2003} are able to distinguish between dynamically cold velocity dispersions associated with regions of star formation, and higher velocity dispersions associated with AGN and Low Ionization (Nuclear) Emission-line Regions (LI(N)ERs), the latter of which is often associated with an aging stellar population \citep{Fernandes2009a}, or with shocked gas \citep{Sanchez2018}. 

Recent work has also demonstrated the importance of considering the impact of diffuse ionized gas (DIG) in spectral measurements \citep[e.g.,][]{Zhang2017, Poetrodjojo2019}. DIG is thought to be a warm component of the ISM, distinct from \HII~regions both by its much more diffuse nature, and its emission line ratios, which are aligned with a harder ionizing radiation source. It has been suggested that a powering mechanism for the DIG light may be an aging stellar population \citep{Zhang2017}, and DIG light and emission from \HII~regions may be spatially coincident. Strong DIG contamination has the effect of pushing a given spectrum towards the LINER region of a BPT diagram, effectively inflating the \NII$_{6584}$/H$\alpha$ line ratio, and thus serving as another potential source of bias to metallicity measurements.

While the spatially resolved data of an IFS survey means we may calculate metallicities for a portion of a galaxy even if the nuclear region is dominated by non-stellar light, previously impossible with single-fibre data, we now also have the potential for unusual behaviour at the \emph{boundary} between those spaxels identified as driven by photoionization by young stars, and those spaxels identified as powered by some other mechanism. BPT classifications are applied to the data in a binary fashion, but physically we would expect a gradual decrease in hard ionization power, described by an inverse square law. It is possible that our BPT diagrams (and other similar diagnostic diagrams) place sufficient constraints on the ionization state of the spaxels in IFS surveys that the metallicities calculated based on emission line ratios in spaxels immediately adjacent to a spaxel flagged as non-star forming are unaffected. It is also possible that the existing diagnostics are not sufficiently restrictive, and more conservative cuts are required if the exclusion of hard ionization contamination is important for the science case in question.

In this work, we aim to test this directly, by identifying a sample of spaxels in the final MaNGA Data Release 17 \citep[DR17]{Abdurrouf2022} immediately adjacent to spaxels flagged on the BPT diagram as powered by non-star-forming radiation, and where gas-phase metallicities are calculable. 
This work is organized as followed:
In Section \ref{sec:sample}, we describe our data set, the selection process of identifying metallicities adjacent to spaxels flagged as non-SF, identify how frequently these metallicities are found, and the methodology for calculating metallicity offsets within our sample. In Section \ref{sec:analysis}, we investigate possible sources of scatter and systematic offsets, and in Section \ref{sec:s06} we investigate methods to eliminate affected spaxels. In Section \ref{sec:discussion} we place our results into context, and propose best practices based on our results. Finally, we summarise our findings in Section \ref{sec:conclusions}. We assume a cosmology of H$_0 =70$ km s$^{-1}$ Mpc$^{-1}$, $\Omega_M$ =0.3, and $\Omega_{\Lambda}$=0.7.

\section{Sample selection \& offset calculations}
\label{sec:sample}

We begin with the final data release of the SDSS MaNGA \citep{Bundy2015, Law2015, Yan2016a, Yan2016b, Wake2017}, Data Release 17 \citep[DR17;][]{Abdurrouf2022}, which contains integral field observations of more than 10,000 galaxies in the nearby universe. Galaxies were observed with bundles of between 19 and 127 fibre optic cables \citep{Smee2013} with a typical point spread function of 2.54 arcseconds which corresponds to about 1.8 kpc at $z\sim 0.037$ \citep{Law2016}. The full sample comprises more than 19 million spectra. Typical physical scales for an individual spectral pixel (spaxel), which subtends 0.5 arcsec \citep{Law2016}, are about 0.37 kpc at $z\sim 0.03$. 

Emission line data are taken from the value added {\sc pipe3d} data products \citep{Sanchez2022}, which contains updates in methodology from earlier data releases \citep{Sanchez2016, Sanchez2016b, Sanchez2018}. We repeat the quality assurance steps taken in \citet{Scudder2021} on the DR15 for the reprocessed and expanded DR17 data set. Briefly, we first correct all emission lines for dust extinction, assuming an optically thick Balmer decrement of 2.85, where the signal to noise in the Balmer decrement H$\alpha$/H$\beta$ is S/N > 5.0, and perform a dust correction with an Small Magellanic Cloud (SMC)-like extinction curve \citep{Pei1992}. Dust corrections using a Milky Way (MW)-like dust correction curves were calculated as part of \citep{Scudder2021}, and metallicity values derived from both dust correction methods were found to be broadly consistent with each other. As previous works have used the SMC-like curve, we present metallicities using that correction curve here. 

\subsection{AGN classifications}

In order to both assess where the non-star forming radiation dominated spaxels can be found, and for subsequent metallicity calibrations, we wish to classify all possible spaxels in the DR17 on the \citet[][BPT]{Baldwin1981} diagram. 

We require a signal to noise of $\geq5.0$ in all four emission lines required to place a spaxel on the BPT diagram: \OIII$_{5007}$, \NII$_{6584}$, H$\alpha$ and H$\beta$. In total, 1,746,152 spaxels surpass this S/N cut for classification on the BPT diagram, a factor of 1.8 increase over the DR15 sample with the same S/N cut (954,884 spaxels). 
We simultaneously classify spaxels by the \citet[][S06]{Stasinska2006}, \citet[K03]{Kauffmann2003}, and \citet[K01]{Kewley2001} diagnostic lines. While the S06 diagnostic provides a clean star-forming sample, and the K01 diagnostic line provides a clean AGN sample, for consistency with previous work \citep[e.g.,][]{Teimoorinia2021, Law2021, Scudder2021}, we use the K03 diagnostic in the present work to select our star-forming sample. We determine that 1592413 (91.2 per cent) 
of S/N $\geq5.0$ spaxels fall into the star forming region by the K03 diagnostic criterion, which roughly doubles the star forming sample when compared to the DR15 \citep{Scudder2021}. We note that we  refer to spaxels flagged as non-star-forming on the BPT diagram as `non-SF spaxels' for brevity purposes; a number of physical scenarios may be playing out within any given non-SF spaxel \citep[e.g.,][]{Belfiore2016, Zhang2017, Sanchez2018, Law2021}, which are not limited to the presence of an AGN.

\subsection{Metallicity catalogue}
The parent metallicity sample uses an updated version of the catalogue presented in \citet{Scudder2021}, with the analysis repeated on the DR17. Details of the metallicity calibrations are presented in that work\footnote{A more comprehensive presentation of the DR17 metallicity sample will be presented in forthcoming work (Scudder et al., in prep).}. For the present work, we use a set of 6 metallicities calculated according to their original publication methods. For the metallicity calibrations themselves, we require S/N $> 5$ in all required lines. We also exclude any spaxels which have H$\alpha$ equivalent widths of less than 6\AA, and more than 1000\AA, consistent with both \citet{Scudder2021} and \citet{CidFernandes2011}, which ought to exclude aged stellar populations \citep{Sanchez2018}.  Metallicities are calculated for all possible spaxels, resulting in a sample of more than 1.5 million metallicity values per calibration. This is an increase of 500,000 gas phase metallicities over the DR15 sample, which was at the time the largest known catalogue of gas-phase metallicities. 

Within this set of 6 calibrations, there are four sets of fundamental emission lines, which are then benchmarked to a metallicity either through theoretical models, or via a set of local observations, and which result in some polynomial form to translate between strong emission line ratios and the resultant log(O/H) + 12 metallicity value. Our sample of 6 metallicity calibrations samples one metallicity for each of three distinct emission line ratios, plus three calibrations which use the N2 line ratio. The results of \citet{Scudder2021} indicate that calibrations using the same emission line ratios generally convert into each other with extremely minimal scatter, so we can expect the same general behaviour among metallicities with the same underlying emission line ratios.

The first line ratio is R$_{23}$ \citep{Pagel1979}, which is defined as: 

\begin{equation}
R_{23}=\frac{\text{\OII}_{3727}+\text{\OIII}_{4959,5007}}{\mathrm{H}\beta}.
\end{equation}

R$_{23}$ is double-valued, and so to distinguish between the high metallicity solution and the low metallicity solution, the \NII$_{6584}$/\OII$_{3727}$~ratio is used to break the degeneracy. A break point of log(\NII$_{6584}$/\OII$_{3727}$) $> -1.2$ marks the divide between the `upper branch' and `lower branch' solutions. 
We select the \citet[KE08]{Kewley2008} recalibration of the \citet{Kewley2002} calibration as the $R_{23}$ representative for the present work. 

Second, there is the O3N2 line ratio, which was first introduced by \citet{Alloin1979}. 
\begin{equation}
\mathrm{O3N2} = \frac{\text{\OIII}_{5007}/\mathrm{H}\beta}{\text{\NII}_{6584}/ \mathrm{H}\alpha}.
\end{equation}
In the present work, we select the \citet{PP04} calibration to represent the O3N2 calibrations (henceforth PP04 O3N2). 

Third, there is the simplest line ratio, the N2 line ratio \citep{StorchiBergmann1994}, which uses only two lines: \NII$_{6584}$/H$\alpha$. We use three N2 calibrations in the present work; PP04 N2, \citet[M13]{Marino2013} N2, and \citet[C17]{Curti2017} N2, for reasons we will explain in detail in Section \ref{sec:offsets}.

Finally, there is the N2S2 line ratio, which is used by \citet[D16]{Dopita2016} alone. This calibration requires both \NII$_{6584}$/H$\alpha$ and \NII$_{6584}$/\SII$_{6717+6731}$. As this is the only metallicity calibration using these line ratios, we include the D16 calibration in the present work. 

Metallicity calibrations are calibrated either to theoretical photoionization models such as \citet{Kewley2002}, {\sc{mappings}} III \citep{Sutherland2013}, or {\sc{mappings}} V \citep{Sutherland2018}, which are used for both KE08 and D16 metallicities to translate between photoionization state and abundances to strong line emission line ratios. By contrast, PP04, M13 and C17 benchmark their line ratios to local, stacked observations of partially resolved galaxies where direct measurements of the electron temperature of the gas are possible, comparing to the \OIII$_{5007}$ line to detections of the traditionally weak \OIII$_{4363}$ line.

Our final sample thus contains metallicities calculated via: KE08, D16, PP04 O3N2, PP04 N2, M13 N2, and C17 N2. The full number of spaxels with metallicities calculated by these six calibrations in the DR17 is shown in the leftmost column of Table \ref{tab:sample}.

 \begin{table*}
	\centering
	\caption{For each metallicity calibration, we show the total number of spaxels with calculable metallicities, and the number of galaxies represented in that number of spaxels in the first and second columns respectively. In the third column we present the number of metallicity spaxels which are identified as bordering a non-SF spaxel, and in the fourth, the percent of the total spaxel count that sample represents. In the fifth column we show the number of galaxies represented in our border spaxel sample, and in the 6th the percent of the number of metallicity hosting galaxies which have metallicities in our border spaxel sample. In the 7th and 8th columns, we show the mean and median number of border spaxels per galaxy.}
	\label{tab:sample}
	\begin{tabular}{lcccc c c c c }
\hline 
Metallicity &  Total &  Total& Border & \% total & Galaxies with &  \% total & Mean  & Median  \\
calibration &  spaxels  &  galaxies & spaxels & spaxels & border spaxels & galaxies & spax/galaxy &spax/galaxy\\
\hline 
KE08 & 1,569,559 & 5,219 & 21,528 & 1.4\% & 1,208 & 23.1\% & 17.8 & 9 \\
D16 & 1,565,452 & 5,186 & 21,046 & 1.3\% & 1,187 & 22.9\% & 17.7 & 9 \\
PP04 O3N2 & 1,590,222 & 5,255 & 21,638 & 1.4\% & 1,209 & 23.0\% & 17.9 & 9 \\
PP04 N2 & 1,586,127 & 5,253 & 21,244 & 1.3\% & 1,203 & 22.9\% & 17.7 & 9 \\
M13 N2 & 1,587,482 & 5,255 & 21,639 & 1.4\% & 1,210 & 23.0\% & 17.9 & 9 \\
C17 N2 & 1,574,395 & 5,231 & 18,326 & 1.2\% & 1,155 & 22.1\% & 15.9 & 8\\
\hline 
	\end{tabular}
\end{table*}

\subsection{Border spaxel identification}\label{sec:border_ids}
To identify spaxels which are within 1 spaxel of a non-SF flagged spaxel, for each of our six metallicity calibrations, we first identify all galaxies in MaNGA with at least one metallicity value. This results in a sample of about 5,200 galaxies, with the exact values shown per calibration in the second column in Table \ref{tab:sample}. This single-spaxel boundary region is chosen as the smallest unit data point present in the MaNGA data, as it corresponds to the area over which BPT and similar classifications can be undertaken. The physical scale of the 0.5 arcsec spaxel will correspond to 0.35 kpc at a typical redshift of 0.03.

\begin{figure}
 	\includegraphics[width=\columnwidth]{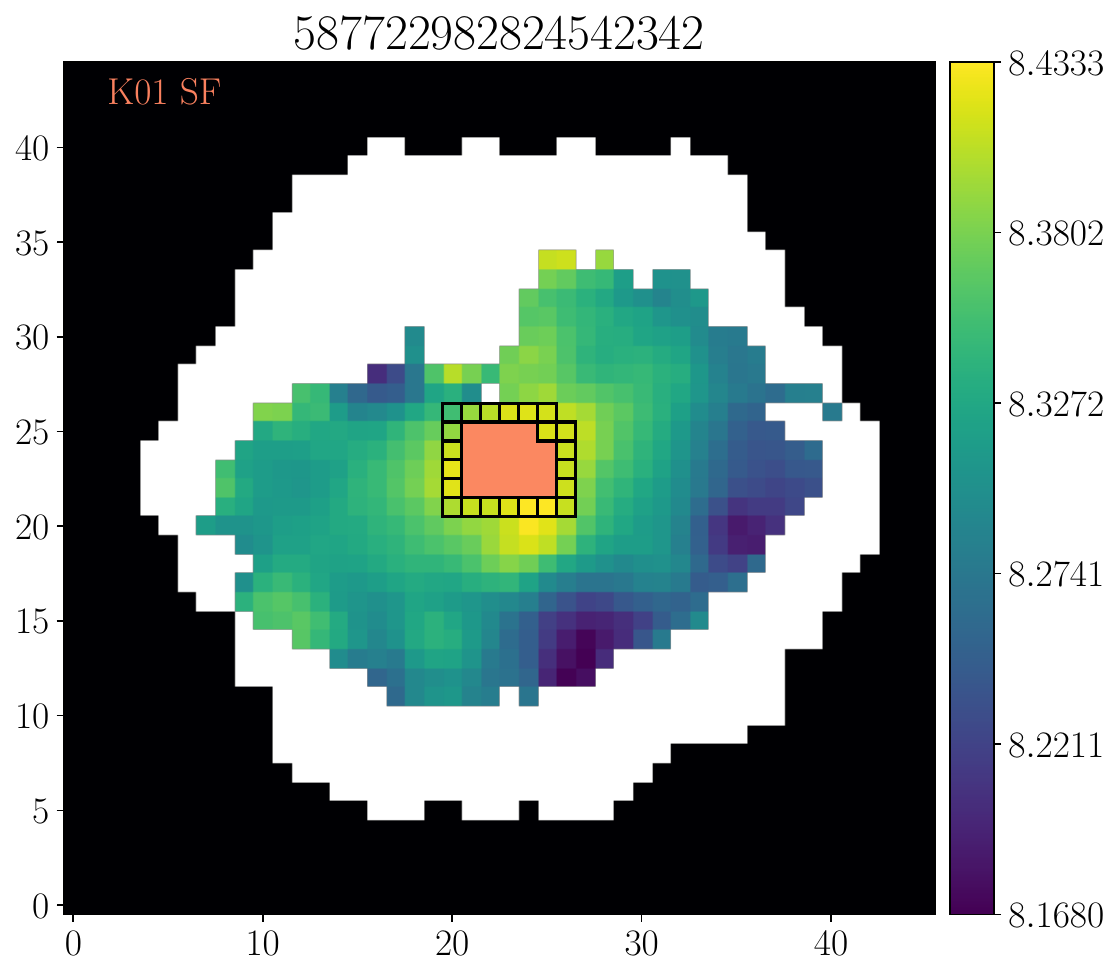}
 	\caption{An example galaxy, with objid labeled at the top of the image. The purple to yellow colour bar indicates the metallicity, calculated here using the PP04 N2 metallicity calibration. In orange are the spaxels flagged as non-SF. In the top left corner we indicate where on a BPT diagram these spaxels are found; in this example, the central orange pixels indicate that these spaxels are classified as non-SF by K03, but as star forming by the K01 diagnostic line. Border spaxels are individually outlined in black, and fully surround the central non-SF classified spaxels. Non-SF flagged spaxels with S/N < 5.0 are not displayed.}
        \label{fig:galaxy_example}
\end{figure}

For each of these galaxies, we then select the subset of galaxies which also contain at least one spaxel flagged as non-SF, so that the subset contains galaxies with at least one metallicity spaxel somewhere within the galaxy, and at least one spaxel flagged as non-SF somewhere within the galaxy.
To identify metallicity spaxels which are immediately adjacent to non-SF flagged spaxels, we iterate through all non-SF-flagged spaxels, as they are less numerous than the number of spaxels with metallicities, to identify any non-SF flagged spaxels that lie within one spaxel in $x$ and/or $y$ of a spaxel with a metallicity. For a given non-SF-flagged spaxel, there are a maximum number of adjacent metallicity spaxels of 8; however, isolated non-SF-flagged spaxels are rare, and generally there is a cluster of such flagged spaxels in the central region of the galaxy. This search algorithm will identify any metallicity spaxels which are immediately adjacent to this region, regardless of its shape. It is possible for a spaxel to be identified as adjacent to more than one non-SF flagged spaxel; we retain only the unique spaxels in our final sample. We will henceforth refer to these identified spaxels with metallicities found adjacent to a spaxel flagged as non-SF by K03 as `border spaxels'. For a visual example of border spaxels identified in a galaxy within our sample, we show one such galaxy in Figure \ref{fig:galaxy_example}. In this figure, we show the map of metallicities as calculated by the PP04 N2 metallicity calibration in a dark blue to yellow colormap, with the region of the galaxy flagged as non-SF in orange. The identified border spaxels are each boxed with a black bounding line. 
  
\begin{figure*}
 	\includegraphics[width=\textwidth]{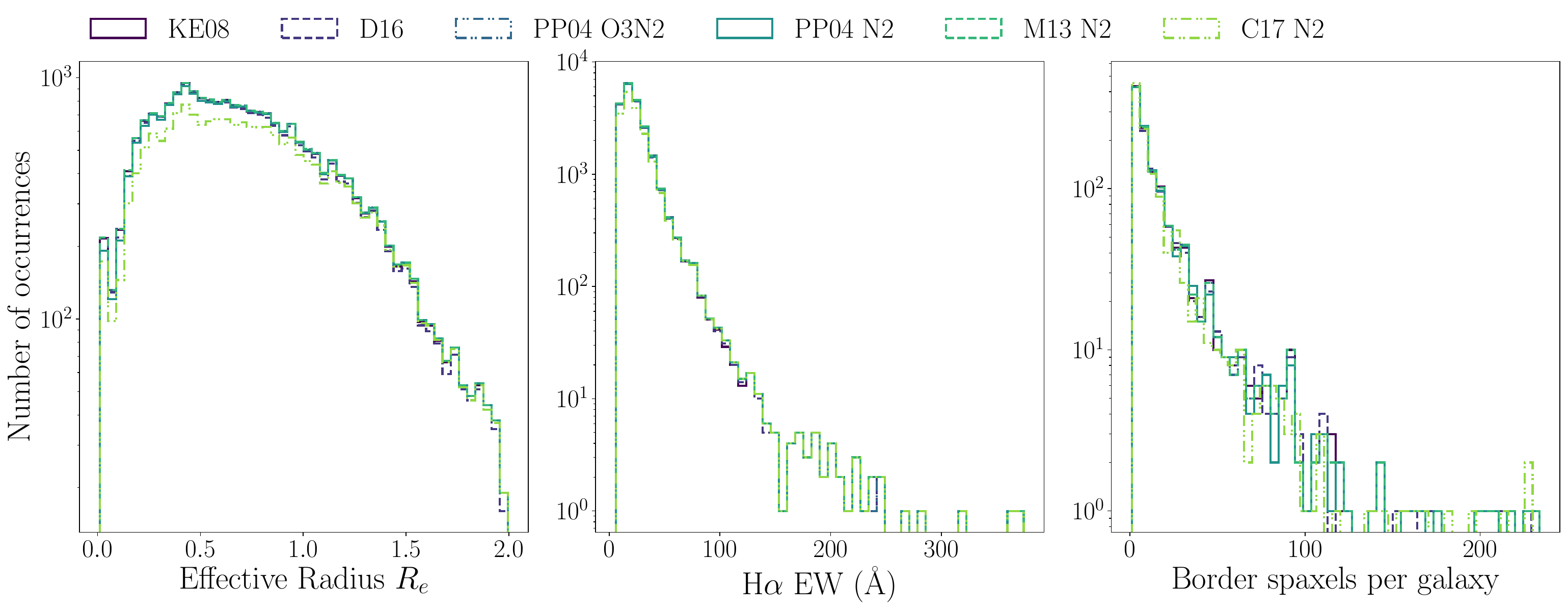}
        \caption{In all panels, our selection of six metallicity methods are colour coded according to the legend above the axes. Left hand panel: Histogram of the distribution of the effective radii of non-SF-bordering spaxels for each of the six metallicity methods shown here. C17 N2 appears to peak at lower values, which corresponds with the smaller sample size for this metallicity (see Table \ref{tab:sample}). Middle panel: Histogram of the H$\alpha$ equivalent widths of all non-SF-bordering spaxels; the distributions are consistent across all metallicity methods. Right hand panel: Histogram of the number of non-SF-bordering spaxels per galaxy, for each of the six metallicity methods shown here, shown in a different colour. As with the middle panel, while there are small numerical differences, the number of border spaxels per galaxy is consistent across metallicity methods.}
        \label{fig:sample_summary}
\end{figure*}

 While our border spaxel identification algorithm is agnostic to position within the galaxy, we additionally constrain that our sample should contain only spaxels within 2$R_e$ of the centre of the galaxy. $R_e$ values use a single component Sersic fit to the r-band presented in the {\sc{pipe3d}} value added catalogue \citep{Sanchez2022} to identify the half light radius. The median bordering spaxel for all samples is found at $r=0.67 R_e$ of the centre, and we show the log-scaled histogram of $R_e$ for all six metallicity calibrations in the left hand panel of Figure \ref{fig:sample_summary}. The lines for each metallicity sample are largely overlaid entirely on each other; the exception being the C17 N2 calibration, which is offset to slightly lower total counts. C17 N2 has a lower overall sample size (see also Table \ref{tab:sample}), and the peak of the distribution is not offset with respect to the other five metallicity samples. We explore this reduction in sample size in more detail in Section \ref{sec:s06};  briefly, this is due to many of the C17 N2 border spaxels exceeding the upper limit in \NII/H$\alpha$ where the calibration is valid. We have also repeated the analysis which follows with a less restrictive ($R_e < 4$) cut, and found no significant changes to our results.   
 
The distribution of H$\alpha$ equivalent widths for our final sample is shown in the middle panel of Figure \ref{fig:sample_summary}, with a median value across all metallicity calibrations of $\sim$ 20.7\AA. 
 Here, as with the $R_e$ distributions, the only metallicity which is flagged as inconsistent with the others is C17 N2, with a minimum $p-$val excluding the null hypothesis at $<4\sigma$; the other 5 metallicities are all consistent with each other, with a minimum $p-$val$=0.9986$, or $\ll 0.5\sigma$ confidence of excluding the null hypothesis that the populations are drawn from the same parent population.
  We discuss the impact of the C17 N2 calibration being slightly less well matched to the remaining five metallicity calibrations, as well as potential sources of this difference, in Section \ref{sec:s06}.
  
  In the right hand panel of Figure \ref{fig:sample_summary}, we show the number of non-SF-bordering spaxels identified per galaxy in our samples.  Across all metallicity calibrations, the median value is between 8-9 spaxels per galaxy, though the mean is substantially higher, at $\sim 18$ spaxels per galaxy, with exact values reported in the second and third from left columns of Table \ref{tab:sample}. 
   As the right hand panel of Figure \ref{fig:sample_summary} is extremely non-gaussian in shape, it is not surprising that the mean and median are not well aligned; the mean will be reflecting the small number of galaxies seen to have very large numbers of bordering spaxels, while the median will reflect the more common (and smaller) number of bordering spaxels per galaxy. The right hand panel of Figure \ref{fig:sample_summary} also shows the consistency in the number of bordering spaxels per galaxy, with small number statistic variations appearing only at large numbers of border spaxels per galaxy. The smallest $p-$value reported for a KS test among all permutations of the metallicity samples is 0.138 (between PP04 O3N2 \& C17 N2), an exclusion of the null hypothesis at $<1.5\sigma$ ability to exclude the null hypothesis of being drawn from the same parent population; the vast majority of combinations report $p-$vals of $\gg0.99$, which is unable to exclude the null hypothesis at any level of significance. 

Our final samples of non-SF-bordering spaxels are summarised in Table \ref{tab:sample}. For each of the six metallicity calibrations (first column), we identify a sample of about 21,000 spaxels with metallicities found within 1 spaxel's distance of a spaxel flagged as non-SF (4th column), in a population of around 1,200 unique galaxies (6th column). Across all six samples, these 21,000 spaxels represent approximately 1.3 per cent (fourth column) of their respective full metallicity catalogues (second column). 
 As the final MaNGA data release comprises some 10,000 galaxies, non-SF-bordering spaxels are found in approximately one tenth of the overall sample of galaxies. However, the fraction of galaxies which have calculable metallicities is substantially lower, at $\sim$5200 galaxies (second column of Table \ref{tab:sample}), so the fraction of galaxies with calculable metallicities with at least one metallicity spaxel which immediately borders a non-SF-flagged spaxel is higher, at around 23\%, with exact values and percentages shown in the 5th and 6th columns of Table \ref{tab:sample} respectively.

\subsection{Placement on diagnostic diagrams} 
\begin{figure*}
	\includegraphics[width=0.95\textwidth]{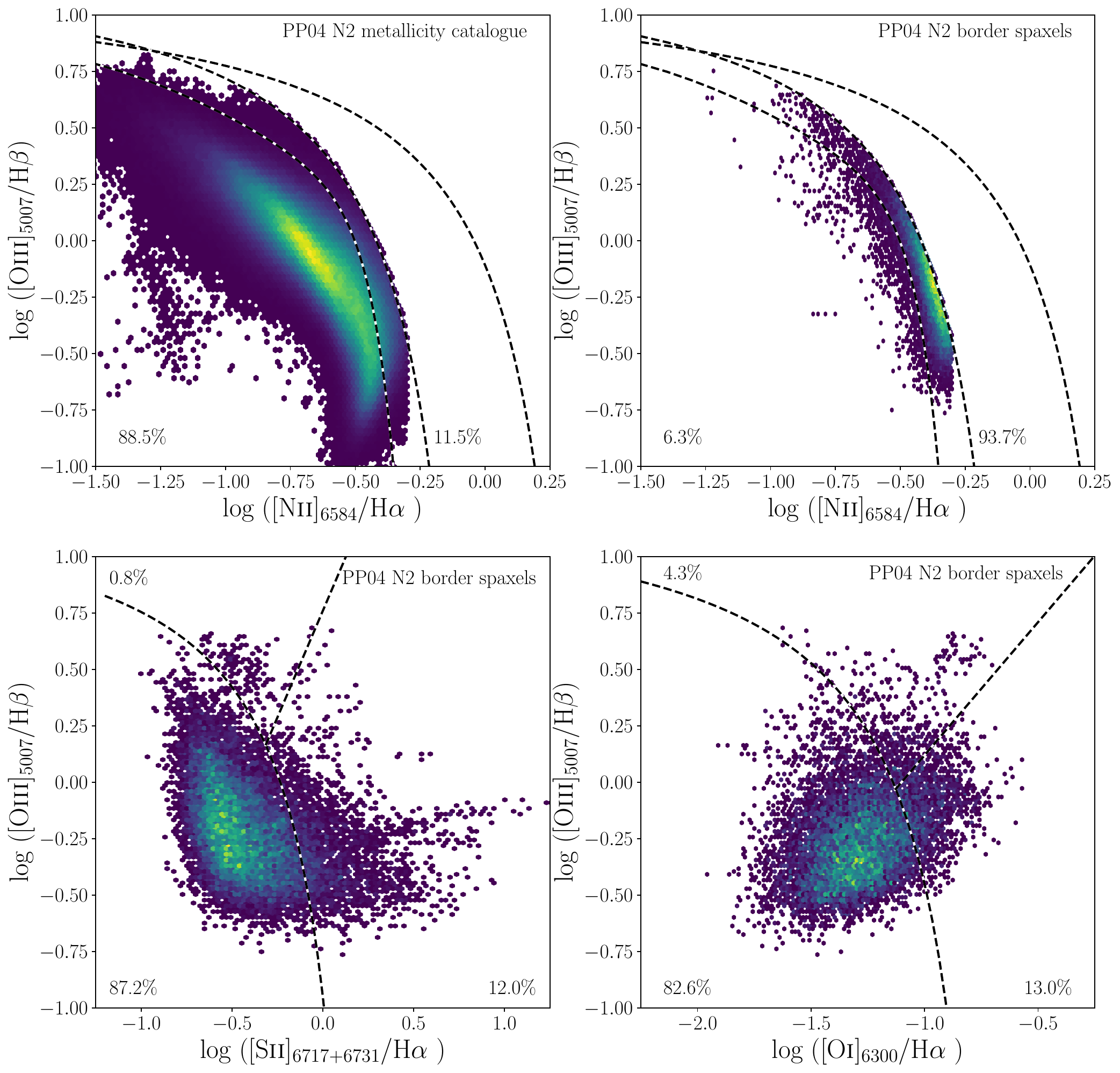}
    \caption{Density histogram of the BPT placement of spaxels with PP04 N2 metallicities. In all panels, colours indicate the log of number density. and in black and white dashed lines, we indicate  diagnostic lines; in the upper panels, the leftmost line is that of S06, the centre line, K03, and the rightmost line is K01, for reference only, since the full metallicity sample uses the K03 line as part of the sample construction. Upper left hand panel: the complete metallicity sample of 1.59 million metallicities calculated by the PP04 N2 calibration and their placement on the BPT diagram. We note in the bottom left corner the fraction of the sample which is found to the left of S06 (SF classification; 88.5 per cent), and in the bottom right corner the fraction classified as SF by K03 and non-SF by S06 (11.5 per cent). Upper right hand panel: The same density histogram but for non-SF-bordering spaxels only. 93.7 per cent of spaxels are found in between the K03 (centre) and S06 (left) curves, with only 6.3 per cent classified as SF by the S06 diagnostic curve. In the lower left panel, we show the \SII~diagnostic diagram for the PP04 N2 border spaxels; to the left of the curved line is the SF region, which holds 87.2 per cent of spaxels. The upper wedge of the diagram is for Seyfert AGN, which comprises 0.8 per cent of the sample, and the remaining right hand side of the diagram flags LINERs (12 per cent). In the lower right hand panel, we show the \OI~diagnostic diagram. Regions marked by black and white lines mark the same regions as the lower left hand panel. In this diagnostic, 82.6 per cent of the PP04 N2 border spaxels are flagged as SF, 13.0 per cent identified as LINERs, and 4.3 per cent as Seyfert AGN.}
    \label{fig:bpt_diagram}
\end{figure*} 

As an initial assessment of the border spaxels, in the upper panels of Figure \ref{fig:bpt_diagram}, we show the log scaled density histograms of spaxels on the traditional BPT diagram, for the PP04 N2 calibration as an example. In the upper left hand panel of Figure \ref{fig:bpt_diagram}, we show the full sample of all spaxels in MaNGA with a calculable PP04 N2 metallicity (1.59 million spaxels). We mark in the bottom left of the panel the fraction of those spaxels which are flagged as star forming via the S06 diagnostic curve (88.5 per cent), and the fraction found in between the K03 and S06 curves (11.5 per cent). We also display the K01 diagnostic line as a point of reference. No metallicities are found in this region of the diagram, since the K03 diagnostic is part of the definition of our metallicity sample.

The upper right hand panel shows the placement of our spaxels bordering non-SF spaxels on the BPT diagram. Here, the population trends are more than reversed; only 6.3 per cent of the bordering spaxels are classified as star forming by S06, while 93.7 per cent are found between the K03 and S06 diagnostic lines. We also note that the vast majority of these bordering spaxels are found clustered very close to the K03 line, dropping in density as \NII$_{6584}$/H$\alpha$ decreases, a trend which is not visible in the left hand panel.
  
In the lower panels, we show the placement of these border-identified spaxels on two alternate diagnostic diagrams, which use the \SII~and \OI~lines in lieu of the \NII~line.

\subsubsection{The \SII~diagnostic diagram}
We impose a detection S/N $>5.0$ requirement for the \SII$_{6717}$ and \SII$_{6731}$ lines separately in order for a given spaxel to be considered classifiable on the \SII~diagram.  This doublet is relatively strong, and so imposing its detection (S/N $>5$) does not result in a substantial reduction in sample size; about 1600 spaxels (7.6 per cent) are lost. We show the exact number of classifiable spaxels per metallicity calibration in the second from left column of Table \ref{tab:bptclass}. For comparison, the leftmost column of Table \ref{tab:bptclass} shows the full sample as selected with a K03 BPT classification. 
 
 Our sample of non-SF bordering spaxels is largely classified as star forming by the \SII~diagnostic diagram; $\sim$87 per cent of the border spaxel sample is classified as star forming across all metallicity methods. In the lower left panel of Figure \ref{fig:bpt_diagram}, we show the \SII~diagnostic diagram for the PP04 N2 sample of border spaxels, where 87.2 per cent of the sample is classified as SF, with 12.0 per cent classified as a LINER, and 0.8 per cent classified as an AGN. In the third column of Table \ref{tab:bptclass}, we report the number of spaxels which are classified as star forming by both the BPT K03 diagnostic and the \SII~diagnostic for all metallicity calibrations; in general, this is about 17,000 spaxels. 
In Appendix Figure \ref{fig:SII_diagnostic}, we show the \SII~diagram for all six metallicity calibrations as a density histogram for all classifiable (S/N > 5.0 ) border spaxels, where we also note the exact percentages of each metallicity sample falling into SF, LINER, and Seyfert portions of the diagram.
 
\subsubsection{The \OI~diagnostic diagram}
 
\OI$_{6300}$ is moderately weaker than \SII$_{6717}$ and \SII$_{6731}$, so we reduce our threshold for a detection to S/N $>3$. \OI$_{6300}$~permits placement on the \OI~diagnostic diagram for approximately 12,000 spaxels, or a reduction of about 9,000 spaxels (20 per cent) when compared to the parent sample. As with the \SII~diagnostic diagram, exact values of the classifiable sample, per metallicity calibration, are reported in the 4th column from left in Table \ref{tab:bptclass}.
 
Similar to the results of the \SII~diagnostic diagram, non-SF bordering spaxels are largely classified as star forming on the \OI~diagnostic diagram. In the lower right panel of Figure \ref{fig:bpt_diagram}, we show the \OI~diagnostic diagram for the border spaxels with PP04 N2 based metallicities. In the \OI~diagnostic, 82.6 per cent of the sample is classified as SF, 13.0 per cent as LINER, and 4.3 per cent as an AGN. These percentages are consistent across all metallicity calibrations, and we show the diagnostic diagram for each metallicity calibration as a density histogram in Appendix Figure \ref{fig:OI_diagnostic}, where we also report the exact percent of each metallicity sample that falls into which regions of parameter space, for all six metallicity calibrations.

\subsection{Control sample \& metallicity offsets in border spaxels}
\label{sec:offsets}
To determine whether or not metallicities in non-SF-bordering spaxels are systematically offset by virtue of their adjacency to a non-SF spaxel, we must construct a careful control sample to compare to. With known metallicity dependencies on galactic radius via metallicity gradients \citep[e.g.,][]{Rich2012, Belfiore2017}, total stellar mass in the form of the mass-metallicity relationship \citep[e.g.,][]{Tremonti2004}, resolved stellar mass density \citep[e.g.,][]{Sanchez2013, Barrera-Ballesteros2016, Baker2023a}, local density \citep[e.g.,][]{Ellison2009, Chartab2021}, gas content \citep[e.g.,][]{Moran2012, Barrera-Ballesteros2018}, and redshift \citep[e.g.,][]{Maiolino2019}, this should be done carefully. We test two different methods of constructing such a control sample, which we describe below, and find that our results are qualitatively unchanged regardless of our choice of control matching method. 
\subsubsection{Control Sample Selection}
Our first control sample construction method compares border spaxels to other spaxels with metallicities within the same galaxy, at the same radius, but not adjacent to a non-SF flagged spaxel. This approach allows us to select a control sample from within the same physical system, and has the benefit of making global galaxy properties perfectly consistent between border spaxels and controls. 
For each border spaxel, we identify all other spaxels within the same galaxy, with metallicities, which are not bordering a non-SF-flagged spaxel. For each border spaxel, we select the closest 5 radial matches from the non-bordering spaxels, with a maximum permitted difference in radius of 0.1 kpc. 
In order to ensure that we are not selecting control spaxels at systematically higher radii than the border spaxels, we run a KS test on the radial distributions of the border spaxels and the control sample. We attempt to find the largest sample of controls without the KS test ruling out the null hypothesis. Our limits of 5 controls per border spaxel and the radial tolerance of 0.1 kpc maximise the control sample size while maintaining KS test values that are unable to exclude the null hypothesis. Using 5 controls per border spaxel, all KS tests return $p-$values $>0.99$, which means we are completely unable to rule out the null hypothesis. If less than 5 control spaxels are identified, the spaxel is excluded from the sample.

We construct a second control sample across the full DR17 sample. We identify all unique galaxies which host bordering non-SF spaxels, and find the 50 galaxies without any non-SF flagged spaxels which are most closely matched in total stellar mass, running a KS test on the distribution of stellar masses for both this control pool of purely SF galaxies and the galaxies which host non-SF-bordering metallicity spaxels. We then proceed to match each bordering spaxel to a set of spaxels from the purely SF galaxies in stellar mass surface density ($\Sigma_*$) and in fraction of effective radius ($R/R_e$). For each border spaxel, we identify the 5 best matches in both $\Sigma_*$ \& $R/R_e$, and verify that a KS test on the border and control samples is unable to rule out the null hypothesis. We also require that the difference in $R/R_e$ be less than 0.1. If there are fewer than five control spaxels which match these criteria, we exclude the border spaxel in question. KS tests on our final distributions of $\Sigma_*$ report $p-$vals of 1.0 across all metallicity samples, indicating no statistical power of exclusion. $R/R_e$ shows a lower KS test $p-$val, but range from 0.0065 to 0.06, which corresponds to a $< 3\sigma$ exclusion of the null hypothesis.  
 
We have fully reproduced the work that follows using this second control sample. We find results in qualitative agreement across both control samples; however, as the second method may introduce additional sources of bias by comparing galaxies with different gas masses and in different global environments, we present figures which use the first control sample selection method, which compares spaxels to other spaxels within the same galaxy, and it is the results of this control sample matched sample that we present in Table \ref{tab:sample}.

\subsubsection{Calculation of metallicity offsets}
For each of the border spaxels with metallicities, we subtract the border spaxel metallicity from the median of the control spaxels matched to it.  We use this ``offset'' value ($\Delta$log(O/H)) as our metric of how atypical the metallicities are, if they are found immediately adjacent to a non-SF flagged spaxel, when compared to spaxels that are at the same radius within the same galaxy, but not adjacent to a non-SF flagged spaxel. Positive offsets would indicate a metallicity in the bordering spaxel that is higher than the controls, and negative offsets are indicative of metallicities lower than the control. Any offsets calculated to be approximately zero indicates no systematic shift between the metallicities in the non-SF-bordering spaxels and the median metallicity of the controls.

\begin{figure}
	\includegraphics[width=0.99\columnwidth]{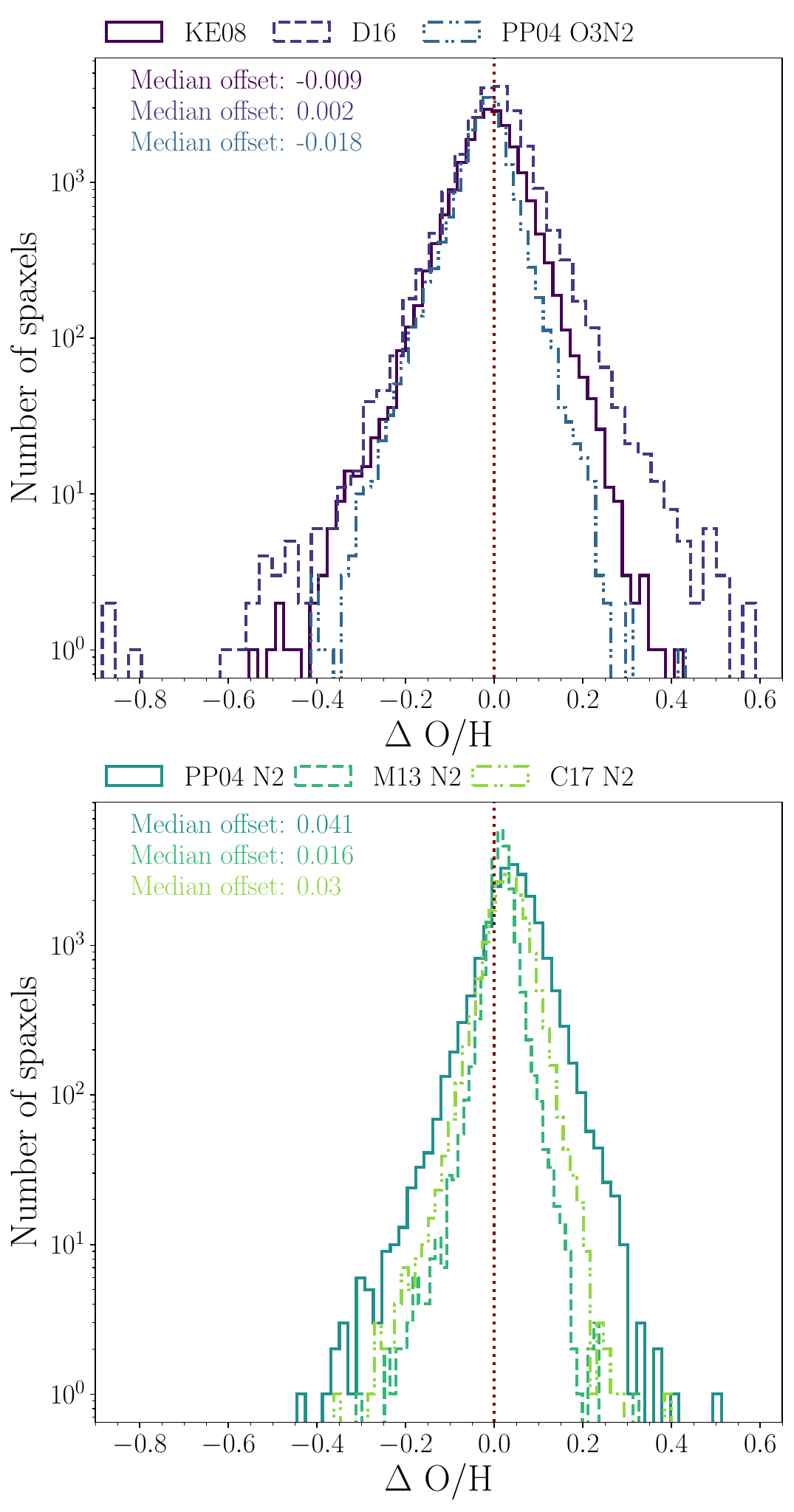}
    \caption{Histogram of the metallicity offsets for non-SF bordering spaxels. Median offsets are labeled in the top corner in the same order and colour as the legend. Top panel: KE08, D16, and PP04 O3N2, in solid dark purple, dashed light purple, and dot dot dashed blue, respectively. While the median offset is very close to zero, there is broad scatter among the offsets of more than a factor of 2.5 in each direction across all three of these metallicity calibrations. Bottom panel: PP04 N2, M13 N2, and C17 N2, in solid teal, dashed green, and dot-dot-dashed pale green respectively. For all three of these calibrations, there is a systematic shift to positive metallicity offsets in the bordering spaxels, compared to a radius-matched control within the same galaxy, ranging from 0.016 dex (M13 N2) to 0.041 dex (PP04 N2). The bottom panel also shows somewhat lower range around the median value.}
    \label{fig:offsets}
\end{figure}

In Figure \ref{fig:offsets}, we show the log scaled histogram of metallicity offsets calculated for each of our six metallicity methods. In the top panel, we show the offsets calculated for KE08, D16, and PP04 O3N2, in solid purple, dashed blue, and dot-dashed pale blue respectively. A vertical dotted dark red line indicates the zero point of no offset from the controls.  Median values for each border spaxel sample are indicated in the top left corner, colour coded to match the line colour, and in the same order as the legend. The three metallicity methods in the top panel are all centred around zero offset, with medians of -0.018 dex (PP04 O3N2), -0.009 (KE08), and 0.002 dex (D16), though the range in offsets per spaxel can be quite broad; spanning between $-0.884$ to $+0.589$ dex (a factor of 30 in range) for D16, between $- 0.553$ to $+0.426$ dex for KE08 and between $-0.414$ to $+0.431$ dex for PP04 O3N2. 

The lower panel of Figure \ref{fig:offsets} shows the set of 3 N2 based calibrations: PP04 N2, M13 N2, and C17 N2, which are shown in teal solid, green dashed, and light green dot-dot-dashed lines respectively. A dark red dashed line indicates the zero point of no offset from the controls. In contrast to the upper panel, in this lower panel, all metallicity methods are offset to higher than expected metallicities, by 0.016 dex (M13 N2), 0.03 dex (C17 N2), and 0.041 dex (PP04 N2). The widest range in individual offsets is PP04 N2, which ranges from $+0.512$ dex down to $-0.445$ dex, followed by C17 N2, with a range of $+0.399$ to $-0.362$ dex, and then by M13 N2 has a range of $+0.326$ to $-0.312$ dex. This median offset to higher metallicity values was initially seen in the PP04 N2 metallicity calibration, and while \citet{Scudder2021} indicated that in general N2-based metallicities trace each other well, we wished to both provide a complete sample and verify this offset to positive values was consistent across this sample of metallicities, and so expanded our initial sample from four to six metallicity methods to be able to include all of the N2 based calibrations.

While all six of our metallicity methods have broad scatter in the total range, extremely offset metallicities are relatively rare, and only those with N2-based calibrations show systematic offsets from the zero point, with the PP04 N2 and C17 N2 calibrations showing the strongest median offsets.

\section{Sources of Scatter \& Bias}
\label{sec:analysis}

We now wish to address two primary questions: whether or not there is an obvious culprit for the broad range of scatter within these samples of non-SF-bordering spaxels, and to identify the source of the bias towards higher offsets within the N2-based metallicities. We further wish to identify whether or not there are straightforward additional constraints or corrections that can be applied to data sets where contamination by harder radiation fields might be particularly important to avoid. 

\subsection{Galaxy-by-galaxy response or general stochasticity?}

We first wish to assess how much of the scatter is being driven by a subset of individual galaxies with highly offset border spaxels, versus a systematic shift of the median values across a large fraction of the sample. For each metallicity sample, we therefore identify each of the $\sim$1,200 galaxies with non-SF-bordering spaxels, and for each galaxy, find the median offset value of all identified border spaxels within that galaxy. 
If we observe a large number of galaxies with median offsets close to zero, then this would tell us that there is no galaxy-wide systematic shift in the non-SF-bordering spaxels. This lack of systematic shift would mean that the range and/or offset in median value is due to individually offset metallicity values, which are atypical for their host galaxy. 
 If, by contrast, the median offsets in a large number of galaxies is shifted to high or low offsets, then the scatter we see in Figure \ref{fig:offsets} is not due to individual stochastic spaxels, but rather a more systematic shift in metallicity values across our entire sample. 

\begin{figure}
	\includegraphics[width=0.99\columnwidth]{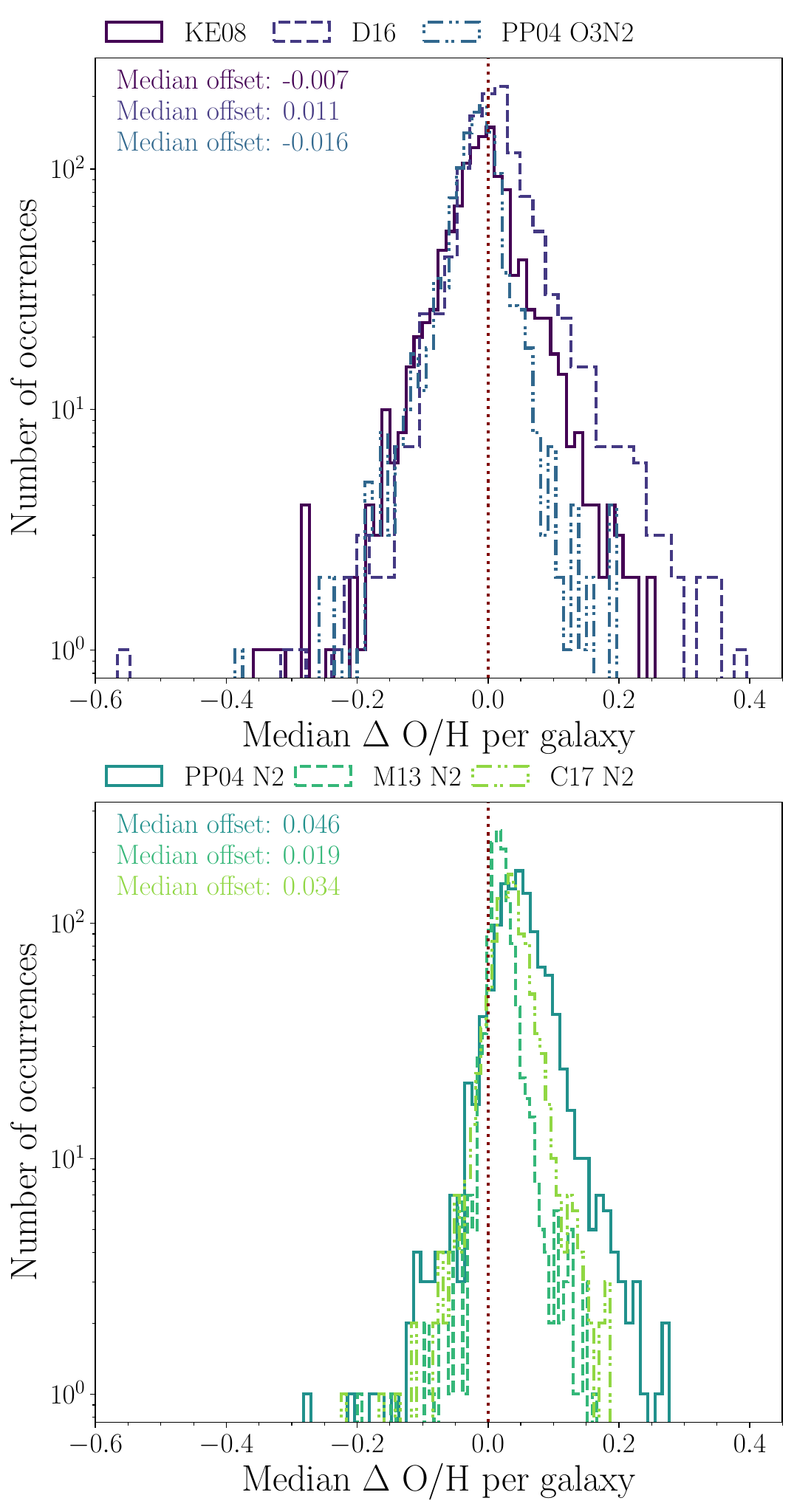}
    \caption{Histogram of the median metallicity offset per galaxy within our samples. In the upper left corner of each panel, we indicate the median offset of our per-galaxy offset distributions, colour coded to match the legend. Top panel, as in Figure \ref{fig:offsets}, shows the offsets for KE08, D16, and PP04 O3N2. These offsets remain centred around zero. Bottom panel shows the same, for PP02 N2, M13 N2, and C17 N2. Typical offsets in the N2-based calibrations remain shifted to positive values, with PP04 N2 the most strongly affected. The zero point is marked by a vertical dark red dotted line in both panels. The range in offsets in both panels is reduced, as we are now looking at median offset values per galaxy; individual outlier spaxels will not be present.  \label{fig:galaxy_offsets}}
\end{figure}

In Figure \ref{fig:galaxy_offsets}, we show the log-scaled histogram of these galaxy-by-galaxy median offsets for all six metallicity methods. In the upper panel, as with Figure \ref{fig:offsets}, we show the distribution of galaxy median offsets for KE08 (solid dark purple), D16 (dashed light purple), and PP04 O3N2 (dot dot dashed blue).
We also indicate the median of these distributions in the upper left hand corner of each panel, colour and order matched to the legend above the axes. The median offsets of each galaxy's offsets are quite similar to the median offset of the overall distribution shown in Figure \ref{fig:offsets}, where the spaxel offsets are presented without being summarised by galaxy. We note that the overall range of offset values has decreased in both panels when compared to Figure \ref{fig:offsets}, which indicates that \emph{very high and low offset values are atypical even for the galaxy they reside within}. However, Figure \ref{fig:galaxy_offsets} indicates that there are still galaxies with median gas-phase metallicities which are offset from their non-bordering controls by 0.1 dex or more. However, these galaxies are in a strong minority, made more visible by the vertical log scaling in Figure \ref{fig:galaxy_offsets}.

In the bottom panel of Figure \ref{fig:galaxy_offsets}, as with Figure \ref{fig:offsets}, we plot the N2 based calibrations PP02 N2 (solid teal), M13 N2 (dashed green), and C17 N2 (dot dot dashed light green), with the median offsets per galaxy marked in the top left corner.  The distributions of these N2 based calibrations remain systematically offset to positive values; the median metallicity offset across all galaxies for PP04 N2 is 0.046 dex higher than the control spaxels in those same galaxies, for M13 N2 the median offset per galaxy is 0.019 dex, and for C17 N2, the median offset is 0.034 dex.

From Figure \ref{fig:galaxy_offsets}, we conclude that some of the range in scatter in Figure \ref{fig:offsets} is driven by individual spaxels in a small number of galaxies. Some of the remaining scatter is likely to be due to a fractionally small number of galaxies which show systematically high or low offsets. However, the shift in the median offset value to positive offsets seen in the N2-based calibration is present across a large number of galaxies, and is not being driven by a subset of atypical galaxies. 

\subsection{Combining multiple diagnostic diagrams}

With individual galaxies ruled out as a source of the broad scatter visible here, we turn to the possibility that these non-SF-bordering spaxels are not truly dominated by star formation, but are identifying ``misclassified'' spaxels on the BPT diagram. Using multiple diagnostic diagrams simultaneously has been suggested as a more reliable way to select ``truly'' star forming spaxels \citep[e.g.,][]{Johnston2023}, removing spectra dominated by aged stellar populations, or shocks, in addition to true AGN. 

\begin{table*}
	\centering
	\caption{Each metallicity method, and how many spaxels are available to be classified via alternate classification diagrams, how many spaxels were flagged as star forming by the \SII~diagnostic, how many were classifiable on the \OI~diagnostic diagram, how many were flagged as star forming by the \OI~diagnostic, and how many passed all three (K03, \SII, and \OI). In the final three columns, we show the number of border spaxels which are flagged as star forming on the stricter S06 BPT diagnostic curve, and then the number of spaxels which are flagged as star forming by both S06 and \SII, and finally those flagged as star forming by S06, \SII, and \OI.}
	\label{tab:bptclass}
	\begin{tabular}{lccccccccccc }
\hline 
Calibration & K03 SF & \SII~class & K03+ \SII~SF & \OI~class & K03 + \OI~SF & K03+\SII+\OI & S06 SF & S06 + \SII & S06 + \SII + \OI \\ \hline 
KE08 & 21,528 & 19,883 & 17,304 & 12,214 & 10,158 & 8,102 & 1,334 & 1,027 & 353 \\ 
D16 & 21,046 & 19,820 & 17,233 & 12,147 & 10,106 & 8,070 & 1,274 & 1,025 & 352 \\ 
PP04 O3N2 & 21,638 & 19,968 & 17,355 & 12,255 & 10,182 & 8,112 & 1,348 & 1,032 & 355 \\ 
PP04 N2 & 21,244 & 19,573 & 17,076 & 11,939 & 9,867 & 7,905 & 1,348 & 1,032 & 355 \\ 
M13 N2 & 21,639 & 19,968 & 17,355 & 12,255 & 10,182 & 8,112 & 1,348 & 1,032 & 355 \\ 
C17 N2 & 18,326 & 16,669 & 14,711 & 9,772 & 7,755 & 6,284 & 1,345 & 1,029 & 352 \\ 
\hline 
	\end{tabular}
\end{table*}

We therefore begin by requiring that a given border spaxel pass both the K03 diagnostic diagram and the \SII~diagnostic diagram. As Figure \ref{fig:bpt_diagram} indicated that approximately 87 per cent of the sample is classified as SF by \SII, this will not substantially reduce the sample size. In Table \ref{tab:bptclass}, we show in the 4th from left column the number of spaxels which pass both K03 and \SII, for all metallicity calibrations, which is a sample of typically $\sim$17,000 spaxels.

Requiring that the border spaxels pass both the BPT K03 criterion and the \OI~diagnostic results in a sample of about 10,000 spaxels (5th column of Table \ref{tab:bptclass}). While the percentage of spaxels which are classifiable on the \OI~diagnostic diagram is about the same as for the \SII~diagnostic diagram, there is an overall reduction in spaxel counts due to the smaller number of spaxels with detectable (S/N > 3.0) \OI~emission lines.

We also test the combination of all three diagnostics by requiring a given spaxel to be simultaneously flagged as star forming by the BPT K03 classification, the \SII~diagnostic, and the \OI~diagnostic. The resulting subsample, shown in the 6th column of Table \ref{tab:bptclass}, consists of around 8,000 spaxels, with the exception of C17 N2, which drops to 6,284 spaxels. Per the work of (e.g.,) \citet{Johnston2023}, spaxels which are triply classified as star forming ought to exclude AGN, shocks, and aged stellar populations.  

\begin{figure}
	\includegraphics[width=\columnwidth]{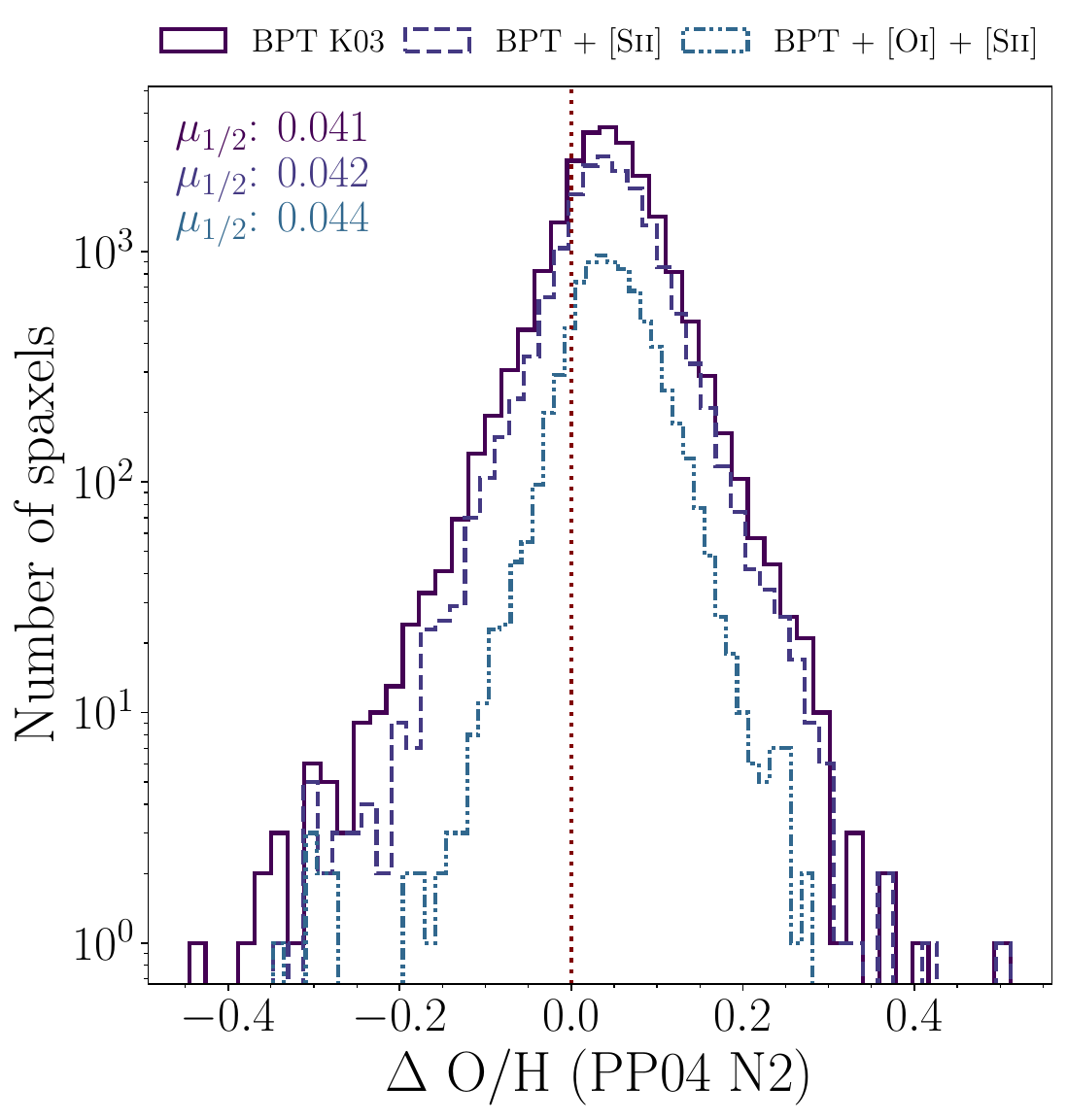}
    \caption{Histogram of the metallicity offsets for PP04 bordering spaxels, with increasing diagnostic requirements. Solid purple histogram shows the offsets if only the K03 BPT diagnostic is required, which represents our full sample, and is the same as the solid teal line in the lower panel of Figure \ref{fig:offsets}. The light purple dashed line shows the distribution of offsets if the \SII~diagnostics must also be passed. Finally, the blue dot-dot-dashed line indicates if the BPT, \SII, and \OI~diagnostic diagrams must all classify a given spaxel as star forming. Medians for all three samples are reported in the upper left corner, with order and colour the same as the legend. Median offsets are consistent across all three samples, as is the range of offsets.}
    \label{fig:hist_bpt_offsets_half}
\end{figure} 

To assess the impact of these additional classifications on both the scatter seen in Figure \ref{fig:offsets} and the systematic shift to positive offsets seen for the N2 based calibrations, in Figure \ref{fig:hist_bpt_offsets_half}, for an exemplar N2 based metallicity (PP04 N2), we plot the metallicity offsets in each of these three subsamples - the full sample delineated by the K03 BPT diagnostic diagram (solid purple line), BPT plus the \SII~star forming criterion (dashed light purple line), and finally, all three (K03 BPT + \SII~+ \OI, dot-dot-dashed blue line). We mark the zero point again with a vertical dotted dark red line. The trends demonstrated here are consistent across all six metallicity samples; PP04 N2 is a representative example.
The median offsets for each of our subsamples with increasing diagnostic classification requirements are listed in the top left corner, in the same order and colour coded identically to the legend. 

We note immediately that the median metallicity offset for non-SF-bordering spaxels is functionally unchanged for all three (sub)samples; while the sample size drops dramatically, the median offset increases very slightly, so these additional cuts are unable to remove the systematic bias to higher metallicities. The difference between the BPT parent sample and the BPT + \SII~ sample is negligible, and while the BPT + \SII~+ \OI~ sample does have a marginally narrower distribution, the range is unchanged. 
This consistency across classification methods indicates that the signal we are looking at is unlikely to be driven simply by a `misclassification' that another diagnostic diagram would have caught.

We therefore find, from experimenting with different BPT diagnostic diagrams, that the inclusion of the \OI~and/or the \SII~diagrams is insufficient to remove the shift to higher metallicities visible in the N2 based metallicities, compared to spaxels which are not immediately adjacent to a non-SF flagged spaxel. The inclusion of these additional diagnostics ought to remove the majority of simple misclassifications from the BPT diagnostic, so our systematic shift is not due to a subsample of misclassified spaxels. Imposing these additional diagnostics do moderately lower the sample size, but only $\sim$12 - 15 per cent of the sample is identified as non-SF on either of the additional diagnostic diagrams. Median offsets are unchanged by the removal of these potentially misclassified spaxels, and so we must find another classification possibility to understand the source of the systematic shift seen in Figure \ref{fig:hist_bpt_offsets_half}.

\section{Placement on the BPT diagram}
\label{sec:s06}

\begin{figure*}
	\includegraphics[width=0.9\textwidth]{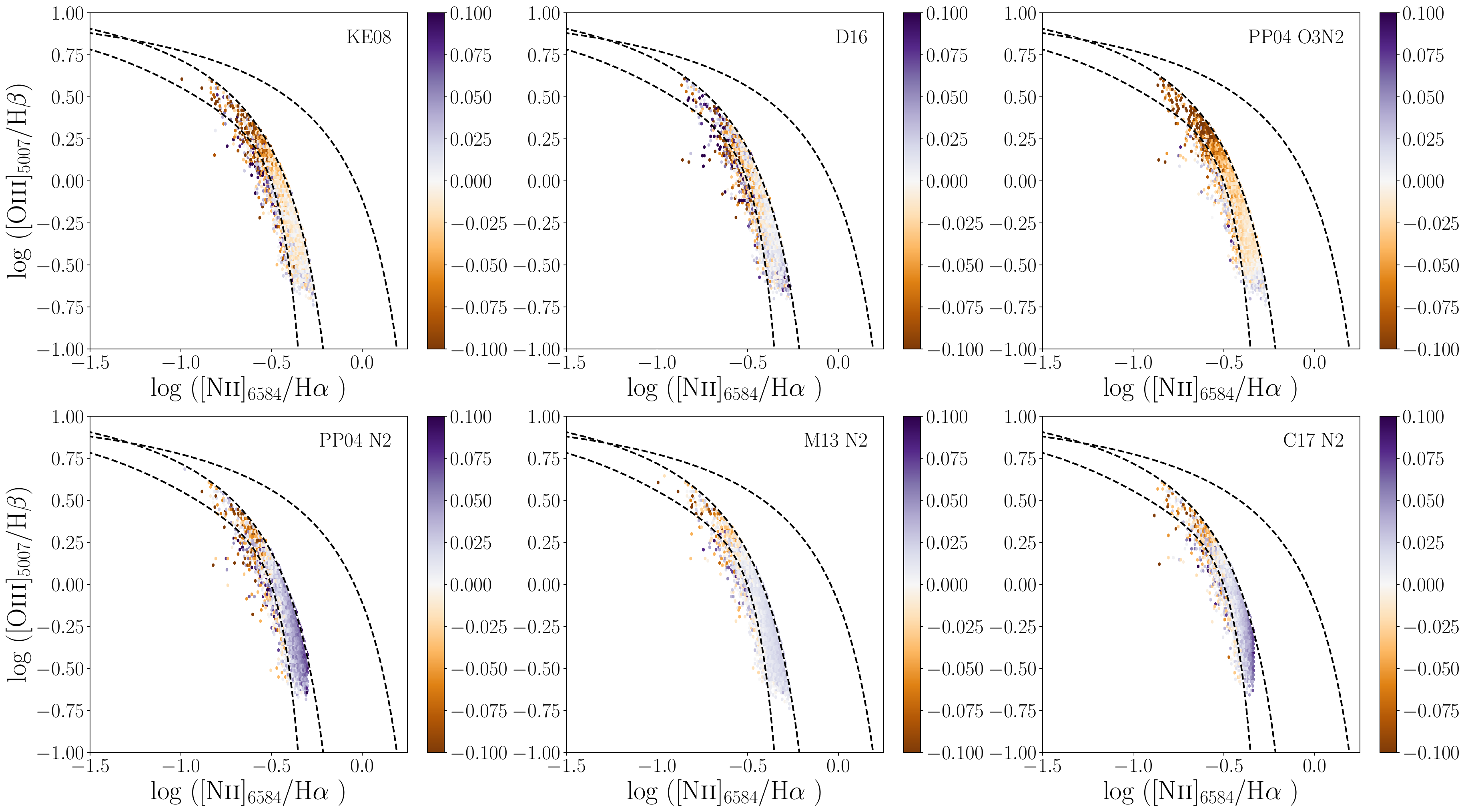}
    \caption{Density histogram of the location of non-SF-bordering spaxels on the traditional BPT diagram, colour coded by median offset value within that bin. Consistent with Figure \ref{fig:bpt_diagram}, we see that all samples are concentrated in the region between the S06 and K03 curves. The PP04 O3N2 calibration seems to trend more significantly to negative offsets with increasing \OIII/H$\beta$ values; however we note from the density in Figure \ref{fig:bpt_diagram} that there are relatively few spaxels in this region. We also note that especially for PP04 N2 and C17 N2, there is a noticeable trend to positive offsets at higher \NII$_{6584}$/H$\alpha$ values. }
    \label{fig:bpt_offsets_diagram}
\end{figure*} 

Since alternative diagnostics themselves are able to remove the systematic offset to higher values in the N2 based calibrations, it is possible that position within the BPT diagram itself is more indicative of likelihood for a non-SF-bordering spaxel to be offset to higher values than non-bordering spaxels. To check whether proximity to the K03 curve is correlated with positive offsets, in Figure \ref{fig:bpt_offsets_diagram} we plot the non-SF bordering spaxels on the BPT diagram in a 2 dimensional histogram, colour coding by the median of the offset values in each bin, with each of the six metallicity samples in its own panel, labeled in the top right corner of the panel. We have fixed the colour bar to be consistent across all panels, and is centred at 0.0. In the top row, where offsets are centred around a median value of zero, we do see some scatter, but no particularly obvious horizontal trends. The PP04 O3N2 calibration trends towards negative offsets at higher values of \OIII$_{5007}$/H$\alpha$, though it seems that the density of spaxels at the highest end of \OIII$_{5007}$/H$\alpha$ is relatively low, as the overall median offset for this sample is still $-0.016$. D16 and KE08 have remarkably flat offset distributions in the BPT parameter space.

In contrast, the lower 3 panels show the N2 based calibrations, where the median metallicity offset is positive. In these panels, especially for the PP04 N2 and C17 N2 calibrations, there is a clear trend to high offsets the closer the data gets to the K03 diagnostic cutoff. M13 N2's offset to higher values is present but as the median offset is smaller than the other two N2 based calibrations, it is less visible when the colour bar scaling is fixed to be consistent across all panels. 

\begin{figure*}
	\includegraphics[width=0.9\textwidth]{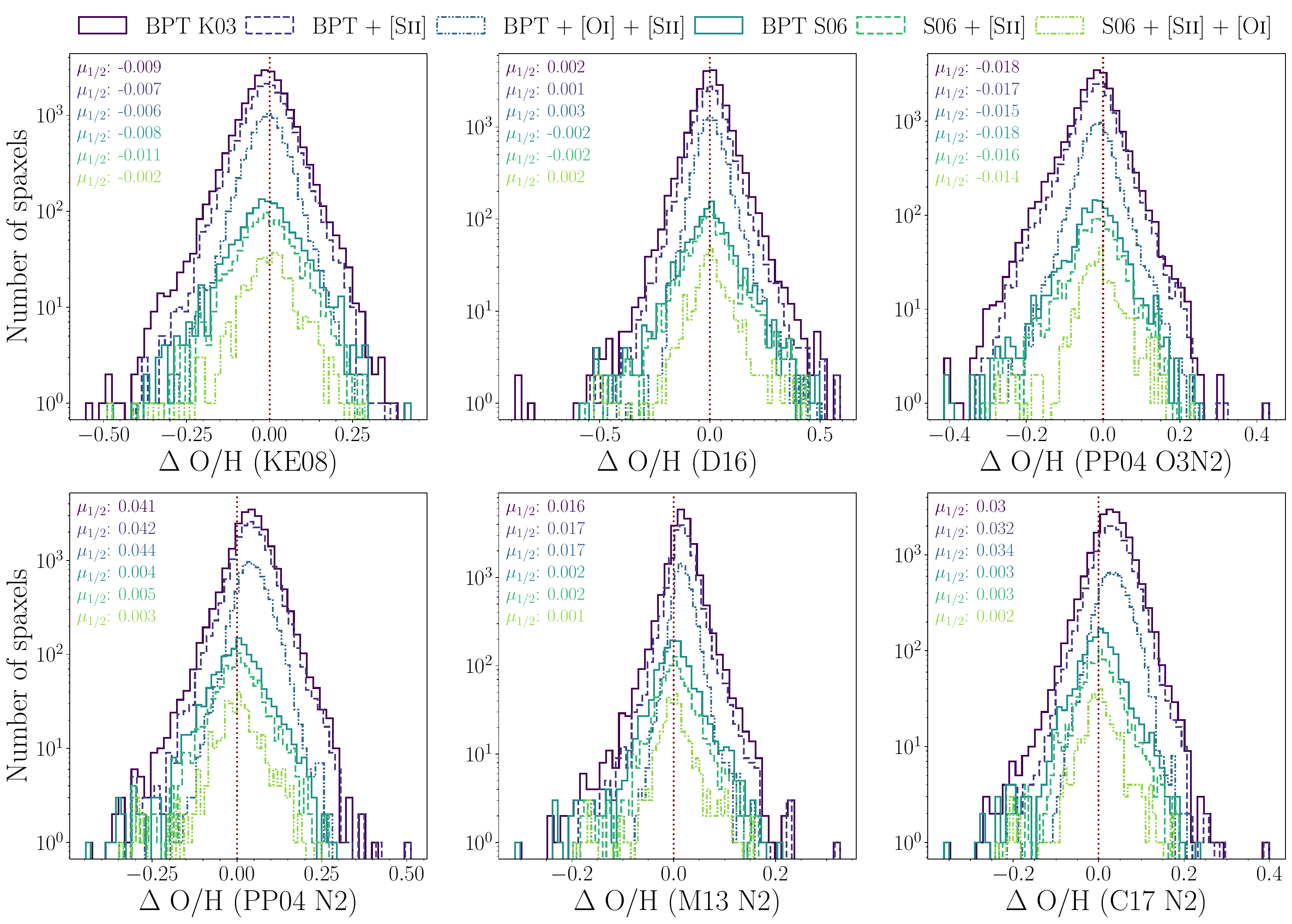}
    \caption{Histogram of the metallicity offsets for bordering spaxels, with all six diagnostic classification subsamples, for each of our metallicity calibrations. The top row shows KE08 (left), D16 (centre), and PP04 O3N2 (right), while the bottom row shows PP04 N2 (left), M13 N2 (centre), and  C17 N2 (right). In all panels, we show the median offsets for each subsample in the top left of each panel, with the colour coding and order the same as in the legend. Imposing a stricter BPT cut does not substantially alter the range of offsets for any of the metallicity methods, nor the median for the top row of metallicity methods. However, it does remove the systematic shift towards higher offsets in the lower row of N2-based metallicity offsets. All panels reflect a roughly order of magnitude drop in sample size with the imposition of the S06 BPT cut.}
    \label{fig:hist_bpt_offsets_all}
\end{figure*}

We note that for both the PP04 N2 and C17 N2 calibrations, metallicities reach the upper limit of their validity within the log(\NII$_{6584}$/H$\alpha$) range that falls between the S06 and K03 lines. The M13 N2 calibration is valid for $-1.6 < $ log(\NII$_{6584}$/H$\alpha$) $<-0.2$, and does not reach its upper limit between the K03 and S06 lines. PP04 N2, by contrast, has a valid range of $-2.5 < $ log(\NII$_{6584}$/H$\alpha$) $<-0.3$, and so there is a visible vertical limit in Figure \ref{fig:bpt_offsets_diagram} where log(\NII$_{6584}$/H$\alpha$) reaches this limit of the calibration. C17 N2 is particularly abruptly cut off, as the upper limit for that calibration is defined as $12+$log(O/H) $< $ 8.85, which corresponds to a log(\NII$_{6584}$/H$\alpha$) limit of around $-0.336$. C17 N2 thus has the most conservative upper limit of the 3 N2 based calibrations. This limit of the calibration's range is also likely to be the reason for the mismatch in the total number of border spaxels for C17 N2 compared to the other five calibrations (e.g., Table \ref{tab:sample}) and the inconsistency in the KS test for the C17 N2 calibration in $R_e$ and H$\alpha$ EW distributions seen in Figure \ref{fig:sample_summary}. 
 In Appendix Figure \ref{fig:n2_elines}, we directly show the correlation between the \NII$_{6584}$/H$\alpha$ emission line ratio and the offset from the non-bordering spaxels for all six metallicity methods. 

Recognizing that Figures \ref{fig:bpt_diagram} \& \ref{fig:bpt_offsets_diagram} indicate that adopting a more strict BPT classification will remove $> 90$ per cent of our bordering spaxels from the sample, we none the less test whether requiring the spaxel be classified as SF by the S06 diagnostic is sufficient to remove the systematic shift towards high offsets in our N2 based metallicity bordering spaxels. We show the resultant number of spaxels passing the S06 BPT classification in the third from right column of Table \ref{tab:bptclass}, which reduces the sample down 1,300 spaxels. If we additionally require the \SII~diagnostic to flag the spaxel as star forming, we drop down to 1,000 spaxels, and with the inclusion of the \OI~diagnostic as well, the sample is reduced to 350 spaxels.   Approximately 33 per cent of the S06 SF sample is classified as SF on the \OI~diagnostic; 58 per cent are lost to non-detections or low S/N, and approximately 3.5 per cent are classified as AGN, and the final 5.5 per cent classified as LINERs. In total, $\sim$ 430 spaxels are classified as SF by both S06 and \OI; so the reduction to 350 spaxels when all three diagnostics are included only includes about 80 spaxels classified as non-SF on the \SII~diagram.

The results of this test are shown in Figure \ref{fig:hist_bpt_offsets_all}, where we plot all six of our diagnostic classification subsamples, for all six metallicity methods. As a point of reference, Figure \ref{fig:hist_bpt_offsets_half} is the same as the darkest three lines in the bottom left hand panel, which shows the values of PP04 N2. We additionally show the S06 cut alone (solid teal line), the S06 + \SII~subsample (dashed green line), and the S06 + \SII~+\OI~subsample (dot-dot-dashed pale green line) in all panels. Median values for each subsample are indicated in the top left corner of each panel.  
For the top three panels, (KE08, D16, and PP04 O3N2) including a stricter BPT diagram cut does not further shift the median value by more than 0.003 dex, though it does, as previously noted, reduce the sample size substantially. We also note that there is not a major reduction in the scatter around the median value, in spite of the drop in sample size.

However, in the bottom panels of Figure \ref{fig:hist_bpt_offsets_all}, the N2 based calibrations, which were previously offset to higher values with the K03 BPT classification, show a reduction in their median offsets by 0.015 to 0.039 dex with the imposition of the S06 criterion. Residual offsets are 0.004 dex (PP04 N2), 0.002 dex (C17 N2), and 0.003 dex (M13 N2). The remaining two additional restrictions (SF by the \SII~diagram and SF by the \OI~diagram) do slightly reduce the median offsets closer to zero, but only by an additional 0.002 - 0.001 dex, consistent with the reduction in median offset from the controls for the top panel figures. Table \ref{tab:bptclass} indicates a consistent drop in sample size between the top and bottom panels; the S06 classification is not excluding \emph{more} spaxels for the N2 based calibrations than it is for the top row of metallicity calibrations. We conclude that the S06 SF classification is the single best additional constraint to remove the enhanced metallicities in border spaxels, though it does come with the collateral function of eliminating over 90 per cent of any such adjacent spaxels from the sample. 

\begin{figure}
	\includegraphics[width=0.9\columnwidth]{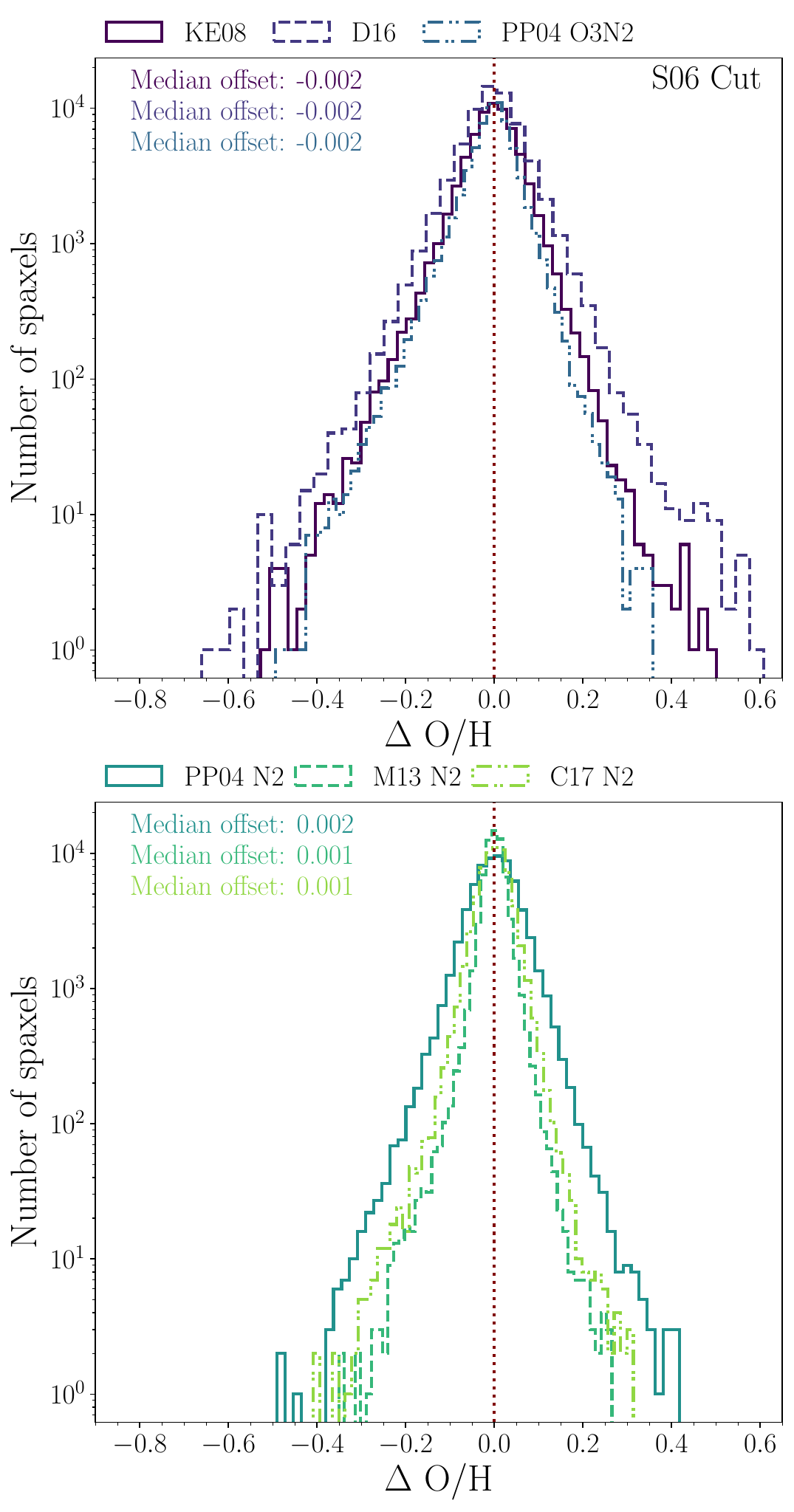}
    \caption{A re-assessment of border spaxels, choosing those identified as SF by S06 and the non-SF sample by S06, rather than K03. Offsets calculated from a  newly defined control sample indicates that the offsets present for K03 flagged non-SF regions disappears when the border is defined as the edges of the S06 classification.}
  \label{fig:S06_border_offsets}
\end{figure} 

\subsection{Alternate selection criteria}
While selecting the S06 subsample of our overall border spaxel sample indicates that the S06 classification is less prone to bias than the K03-based parent sample, the subsample is different from selecting all possible S06-based bordering spaxels from the beginning. We therefore wish to test whether the biases we saw for N2-based calibrations remain if we select only S06 classified border spaxels, or whether - as suggested by our K03 subsampling - the bias disappears. We therefore repeat the identification of border spaxels where metallicities must be immediately adjacent to a non-SF spaxel as flagged by the S06 diagnostic, instead of the K03 BPT diagnostic, outlined in Section \ref{sec:border_ids}. We then construct a new control sample in an identical fashion as before, selecting the 5 closest spaxel matches in radius within the same galaxy, but not identified as bordering an S06-flagged non-SF spaxel. As with the previous sample, if fewer than 5 controls are identified within a maximum permitted difference of 0.1 kpc, the spaxel is excluded from the sample.
The S06-based sample identifies approximately triple the number of non-SF adjacent spaxels: approximately 66,500 across all metallicity calibrations, representing 4.2 per cent of the total metallicity sample, and approximately 40 per cent of the galaxies with metallicities. A complete accounting, in an analogue of Table \ref{tab:sample}, is presented in Appendix Table \ref{tab:s06_sample}.

We then calculate metallicity offsets for all border spaxels with identified control spaxels, and we plot the histograms of these offsets, along with the median value across all border spaxels in Figure \ref{fig:S06_border_offsets}. This figure is therefore an analogue to Figure \ref{fig:offsets}, but created with the stricter S06 cut in the BPT diagram. The upper panel, which displays the offsets for non-N2 based metallicities, is roughly the same as Figure \ref{fig:offsets}. However, the bottom panel has removed all systematic offsets from the N2-based metallicities, consistent with the cuts imposed on the K03-based border spaxel sample shown in Figure \ref{fig:hist_bpt_offsets_all}. We conclude from this final test that the use of the S06 diagnostic line, when used to delineate SF and non-SF spaxels, is effective at removing the systematic bias towards higher metallicities seen in the N2-based calibrations when the K03 BPT diagnostic line is used.

\begin{figure}
	\includegraphics[width=0.9\columnwidth]{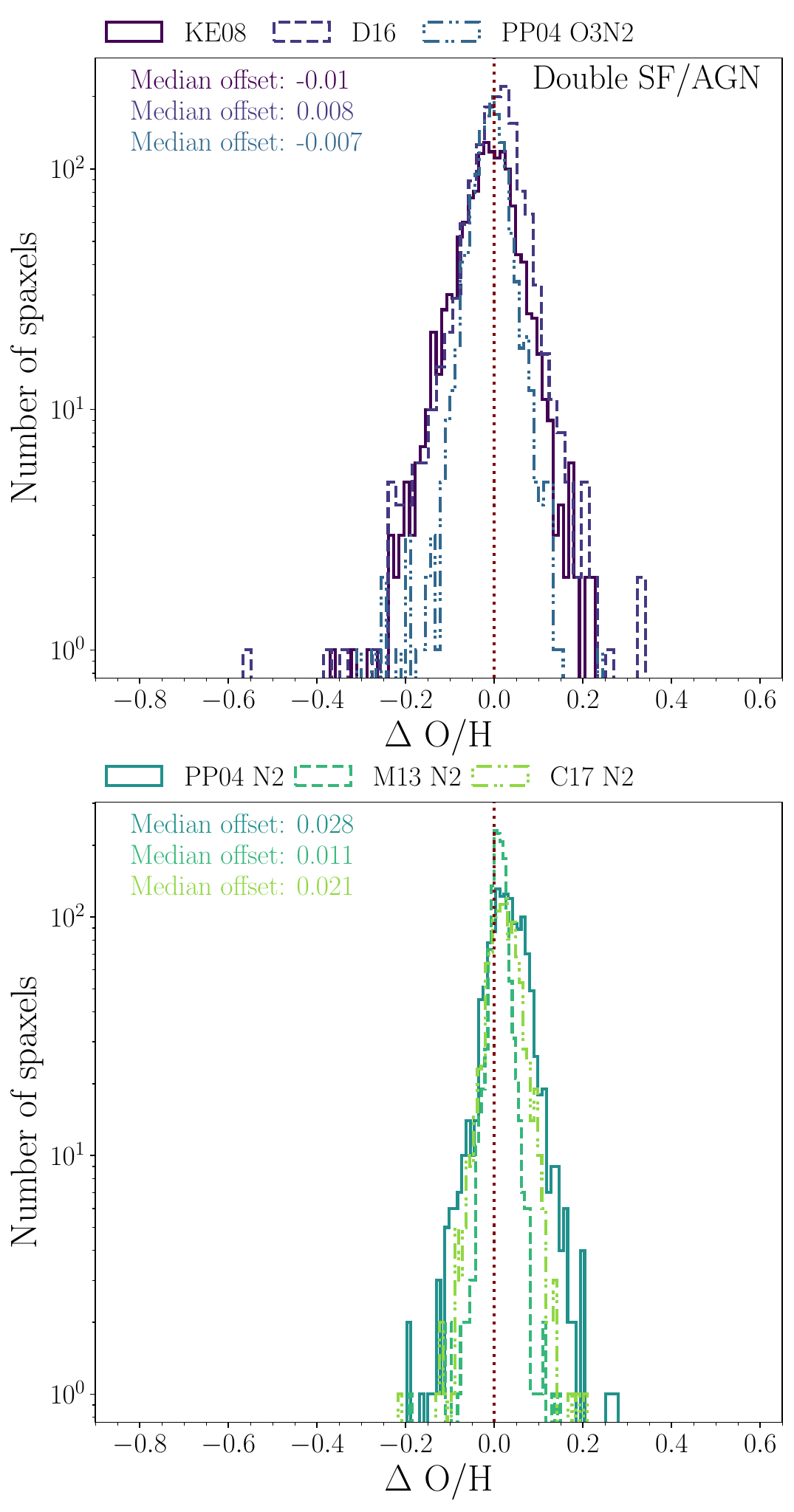}
    \caption{Offset values calculated for those spaxels which are SF by both K03 and the \SII~diagnostic, and requiring the non-SF spaxel which it borders to be non-SF by both K03 and the \SII~diagnostic diagram. We see that while the range is roughly halved, when compared to the fiducial K03 based sample, N2 based metallicity calibrations are still offset by $\sim +0.02$ dex.}
  \label{fig:double_S06_border_offsets}
\end{figure} 

We also run one final additional test, where spaxels are only selected as non-SF if both the K03 BPT diagnostic and the \SII~ diagnostic identify the spaxel as non-SF in nature, and star forming border spaxels are only selected if they are identified as SF by both the K03 BPT classification and the \SII~diagnostic. We require S/N $> 5.0$ for the \SII~diagnostic lines. Spaxels are then matched to a sample of controls within the same galaxies, and the offsets are calculated in an identical fashion as above. The double AGN and double SF criterion results in a much smaller sample of bordering spaxels, with about 1300 spaxels (0.1 per cent of the metallicity sample) identified as bordering a non-SF spaxel. As with the previous test, an analogue of Table \ref{tab:sample} is presented in Appendix Table \ref{tab:doubleSF_sample}.
 
For this sample, shown in Figure \ref{fig:double_S06_border_offsets}, we find that the non-N2 based metallicities have offsets which are broadly consistent with previous samples, with median offsets ranging between -0.008 and +0.1 dex. However, the N2 based metallicities still show systematic offsets in the bordering spaxels. These offsets are of a smaller magnitude than our nominal K03 based sample, ranging from +0.011 to +0.028 dex, but considerably larger than the offsets of +0.002 dex seen for the sample defined with the S06 BPT classification. We do note that this double classification requirement for both non-SF and SF spaxel identification substantially reduces the scatter around the median value. For PP04 N2, offsets range from +0.28 dex to $-0.2$ dex,  from +0.15 dex to $-0.19$ dex for M13 N2, and from +0.21 dex to $-0.22$ dex for C17 N2. These ranges in offset are approximately half of what is seen in the parent sample (e.g., Figure \ref{fig:offsets}).
 
From these additional tests, we confirm that using additional diagnostic diagrams such as \SII, even when requiring the relatively strict requirements that the non-SF spaxel must be identified by two diagnostic diagrams as non-SF, and the bordering spaxel must be identified as SF by two diagnostic diagrams is not sufficient to remove the bias towards higher metallicities in the N2 based metallicity calibrations. The S06 diagnostic line on the BPT diagram, however, is an effective criterion, and without requiring any additional diagnostic diagrams, eliminates the statistical bias to high metallicities.

\section{Discussion}
\label{sec:discussion}

Previous works \citep[e.g.,][]{Belfiore2016, Zhang2017, Law2021, Johnston2023} examining the reliability of applying BPT classifications to IFS surveys started with assessments of whether or not the BPT diagram was still viable when looking at more resolved regions of a galaxy, and in particular if the spectral information from larger galactic radii are still classified as expected. There are two sides of the diagnostic diagram to assess; first, what physical processes the non-SF wing is identifying beyond true AGN, and secondly whether the star forming wing is still a clean sample of regions dominated by young star formation, ionized by O and B stars. The latter is most important for the calculation of metallicities, since the assumption of photoionization underlies the emission line ratios used in these calibrations. However, in the present work, the other side of the diagram is also involved, as it is the potential radiation bleed-through from adjacent spaxels that would be responsible for the systematic offsets we observe in the \NII$_{6584}$/H$\alpha$ based metallicities.  

The non-SF side of the diagram has been suggested to be populated by a mixture of sources of radiation long before the advent of the IFS survey \citep[e.g.,][]{Kewley2006}, and indeed part of the rationale for using \OI~and \SII~diagnostics as alternatives to the traditional BPT diagram is to allow a better disambiguation from Seyfert-style powerful AGN, and LINER-powered sources, which could be low luminosity AGN, gas shocks, or aged stellar populations dominating the spectrum, if there is less active star formation present in a given region \citep[e.g.,][]{Johnston2023}. The placement of the diagnostic curve between SF and non-SF in the BPT diagram attempts to place a reasonable boundary between these populations, but the goals of any individual science case may change which dividing line is best. 

On the whole, the BPT diagram itself has appeared to be robust to the increase in data complexity and the inclusion of spectra from larger radii in IFS surveys, with K03 or similar diagnostic lines, though they were originally designed for single-fibre data \citep[e.g.,][]{K01, Kauffmann2003, Stasinska2006}, proving adequate to select star-forming samples in MaNGA \citep{Law2021}. These existing diagnostics can, for statistical samples, select low velocity dispersion ($\sim 24$ km $s^{-1}$) samples from those with higher velocity dispersions ($< 200$ km $s^{-1}$) typical of AGN or LINERs. \citet{Law2021} examines these trends on the MPL-11 sample, which contains 3.9 million spaxels across 7400 galaxies. Similarly, for the selection of a star-forming sample, while \citet{Johnston2023} recommends the use of more than one diagnostic diagram (as we verify in the present work), so long as the gas shows no signatures of kinematic complexity, and at least two diagnostic diagrams classify the spectra as SF, then it remains classified as a SF spaxel. 
As the vast majority of our identified border spaxels are classified as SF by at least two diagnostics, and some $\sim 75$ per cent of our initial sample are triply classified as SF, it is unlikely that our border spaxels are being misclassified. Even with the triple SF classification, we note that Figure \ref{fig:hist_bpt_offsets_half} still shows a systematic offset in the bordering metallicities. 

However, the spaxels which are flagged as non-star forming on a BPT diagram are much more complex; for the present work, this means that the source of the bleed-through contamination found in our \NII$_{6584}$/H$\alpha$ based metallicities may be coming from multiple sources. Existing work indicates that in addition to the traditional luminous AGN powered by a supermassive black hole (Seyfert AGN), there may be spaxels powered by more obscured or lower-luminosity AGN, which would appear in the LINER regime, aged stellar populations \citep{Belfiore2016}, shocks, and DIG light may all be blended together in the population flagged as ``AGN'' in the BPT diagram. Our selection criteria for the present work is fully agnostic to the radiation source driving any bleed-through; and indeed our goals are primarily to assess any biases in the metallicity catalogue and identify methods to minimize them, regardless of driving source. 

Regardless, it may be of interest to examine potential physical radiation causes for the offsets seen in our \NII$_{6584}$/H$\alpha$ based metallicities. If the spaxels classified as ``AGN'' by the K03 line are also classified as AGN on the \SII~and/or \OI~diagrams, and show large H$\alpha$ EW, then it is likely they are true AGN sources \citep{Johnston2023} and the elevation of the \NII$_{6584}$/H$\alpha$ ratio is naturally explained by the production of a very hard ionizing radiation spectrum by the accreting supermassive black hole. With a higher H$\alpha$ EW, it is likely the culprit is an aged stellar population instead \citep{Belfiore2016, Sanchez2018}. Aged stellar populations may still be producing harder ionization radiation than the photoionization produced by O and B type stars \citep{Zhang2017}, and so our inflation of the \NII$_{6584}$/H$\alpha$ ratio is still due to harder radiation fields. 

A more subtle source of contamination in these spaxels could be the influence of DIG light. While its presence can change emission line ratios as it increases in dominance over any photoionized \HII~regions, the DIG itself requires a powering mechanism, which is potentially at least in part, aged stellar populations \citep{Zhang2017}. Rather than looking at the light from a hot evolved star, the DIG traces the impact of those stars on the ISM. Where DIG light dominates the spectrum, these spectra tend to fall within the LINER regions of diagnostic diagrams \citep{Zhang2017, Byler2019, Johnston2023}, where they can be distinguished from shocked gas primarily by looking for kinematic disturbances in the gas, which are common in shocked gas and less so for DIG light \citep{Johnston2023}, though \citet{Law2021} places the shocked gas along the AGN sequence. 

We therefore select the non-SF spaxels our metallicity sample were identified as being adjacent to, and plot those spaxels on the BPT, \SII, and \OI~diagnostic diagrams to gain a sense of which (if any) of the above radiation sources could be most at play for the metallicity sample. Due to the sample selection of our catalogue, there are no non-SF spaxels to the left of the K03 line in the BPT diagram. The overwhelming majority (99.5 per cent) of the adjacent non-SF spaxels are found to the right of the K03 diagnostic line and to the left of the K01 line, regardless of metallicity calibration. 

As with the metallicity spaxels, the majority of the non-SF flagged spaxels are classifiable on the \SII~diagnostic diagram ($10$ per cent excluded), 
and a larger fraction of them are unclassifiable on the \OI~diagram due to non-detections of the \OI~line ($\sim47$ per cent). 
We show the exact number of non-SF spaxels in Table \ref{tab:agn_breakdown}. The \SII~and \OI~diagnostics classify not just as SF, but as LINER and Seyfert AGN. In our sample of BPT classified non-SF spaxels, the majority of them are classified as star forming on the alternate diagnostic diagrams. Approximately 38 per cent of the total sample is classified as SF on the \OI~diagram ($\sim$ 72 per cent of classifiable spaxels), and 76 per cent are classified as SF on the \SII~diagram (84 per cent of classifiable spaxels). Very few spaxels are classified as a ``true'' (Seyfert) AGN, with less than two hundred spaxels classified as AGN in the \SII~diagnostic, and approximately 500 spaxels (3 per cent) in the \OI~diagram. 

\begin{table*}
	\centering
	\caption{For each metallicity calibration presented here, we indicate the number of spaxels flagged as non-SF which have an adjacent metallicity (second column), and in the third and fourth columns present the number of those spaxels which are flagged as SF by the K01 diagnostic, vs AGN by K01. We then show the number of spaxels classifiable on the \SII~diagnostic diagram (5th column) and the number of those classified as SF (6th column), LINER (7th column), and Seyfert AGN (8th column). Finally, in the rightmost four columns, the number of spaxels classifiable on the \OI~diagnostic diagram (9th column), the number classified as SF (10th column), LINER (11th column), and Seyfert AGN (rightmost column).  }
	\label{tab:agn_breakdown} 
	\begin{tabular}{lccccccccccc}
\hline 
Calibration & BPT class & K01 SF & AGN & [SII] class & [SII] SF & [SII] LINER & [SII] AGN & [OI] class & [OI] SF & [OI] LINER & [OI] AGN \\ 
\hline 
KE08 & 16,923 & 16,864 & 59 & 15,249 & 12,861 & 2,196 & 192 & 9,036 & 6,489 & 2,039 & 508 \\ 
D16 & 16,625 & 16,566 & 59 & 15,148 & 12,780 & 2,172 & 196 & 8,972 & 6,446 & 2,020 & 506 \\ 
PP04 O3N2 & 16,959 & 16,898 & 61 & 15,273 & 12,878 & 2,199 & 196 & 9,044 & 6,489 & 2,041 & 514 \\ 
PP04 N2 & 16,676 & 16,615 & 61 & 14,990 & 12,665 & 2,129 & 196 & 8,842 & 6,289 & 2,039 & 514 \\ 
M13 N2 & 16,960 & 16,899 & 61 & 15,273 & 12,878 & 2,199 & 196 & 9,044 & 6,489 & 2,041 & 514 \\ 
C17 N2 & 14,263 & 14,202 & 61 & 12,602 & 10,732 & 1,674 & 196 & 7,201 & 4,774 & 1,913 & 514 \\ 
\hline 
\end{tabular}
\end{table*}

As the primary goal of this exercise is to attempt to determine what seems like the most likely physical scenario that our adjacent metallicities are proving sensitive to, we examine the subset of these diagnostic diagrams which are in overlap (about 8900 spaxels), presenting a detailed breakdown in Table \ref{tab:disambig}. For each combination of classifications across the three diagrams (K01 SF \& AGN for the BPT diagram, and SF, LINER \& AGN for the \SII~\& \OI~diagrams), we examine how many spaxels fall into that combination of classifications. The largest grouping of identified non-SF spaxels without metallicities are those classified as K01 SF on the BPT diagram, with both of the other diagrams identifying the spaxel as SF (approximately 55 per cent of the sample, across all metallicities). Another $\sim$ 34 per cent of the sample is represented by a K01 SF classification on the BPT diagram, and the other two diagnostics classifying as SF and LINER. Combined with K01 SF + SF + SF classifications, the K01 SF + SF + LINER classifications represent about 90 per cent of the total sample. Triply classified AGN spaxels are very rare, at only $\sim$0.07 per cent of the total sample. Spaxels with two AGN classifications and one SF classification do not exist in this sample. 

Given this distribution of the sample across the possible spaxels, and the placement of these spaxels on the BPT diagram itself, it seems most likely that these non-SF spaxels identified as ``AGN'' by the K03 BPT diagnostic diagram are, in bulk, being affected by the increased impact of aged stellar populations, either directly from viewing the aged stellar light, or indirectly through DIG contamination of the spectra. In either case, our results indicate that the K03 classification on the BPT diagram is quite effective at removing these contaminated spaxels (consistent with the results of \citealt{Law2021}) except in a small number of cases, which themselves can be further removed by using the stricter S06 BPT classification, which will eliminate $\sim$94 per cent of affected metallicities, and $\sim$11 percent of the metallicity sample at large (see Figure \ref{fig:bpt_diagram}).
 
 Regardless of the physical source of our harder ionization source, the results presented here indicate that while such a configuration is unlikely, it is possible for a metallicity to be calculated immediately adjacent to a spaxel flagged as non-star forming in nature. This immediate adjacency does not systematically offset metallicities calculated with R$_{23}$, O3N2, or N2S2 based metallicity calibrations, when compared to metallicities found within the same galaxy at the same radius, but not found bordering a non-SF spaxel. However, it does systematically impact those metallicities which rely solely on the \NII$_{6584}$/H$\alpha$ line ratio. We ascribe this systematic offset, up to 0.041 dex in scale, to the impact of a harder ionization field from the adjacent non-SF spaxel inflating the \NII$_{6584}$~line, and thus the metallicity itself. 
 
 The lack of impact on the other metallicity calibrations is likely due to the presence of other lines in the metallicity calibration. The R$_{23}$ line ratio itself does not actually depend on \NII$_{6584}$~at all; the \NII$_{6584}$~line is used exclusively to determine which of the double-valued solutions should be used. This bifurcation point is at log(\NII$_{6584}$/H$\alpha) = -1.2$, which is far to the left of the vast majority of the border spaxels' values of log(\NII$_{6584}$/H$\alpha$), as shown in Figures \ref{fig:bpt_diagram} \& \ref{fig:bpt_offsets_diagram}. We similarly do not see strong patterns for R$_{23}$ in Figure \ref{fig:bpt_offsets_diagram}. While the O3N2 line ratio does include the N2 line ratio in the denominator, the inclusion of the \OIII/H$\beta$ line ratio in the numerator seems to prevent systematic shifts to positive values. We note that in Figure \ref{fig:bpt_offsets_diagram}, there is a shift to more negative offset values in non-SF bordering spaxels at high \OIII/H$\beta$, but as seen in Figures \ref{fig:offsets} \& \ref{fig:hist_bpt_offsets_all}, this does not seem to result in a substantial shift in the overall distribution of the offset values away from a median value close to 0. 
Finally, the D16 set of line ratios does include the \NII$_{6584}$/H$\alpha$ ratio, but it also includes the line ratio \NII$_{6584}$/\SII$_{6717+6731}$, and the inclusion of this second line ratio does appear to have also stabilised this calibration against direct increases in \NII$_{6584}$~flux. Indeed, the distribution of offsets in Figure \ref{fig:bpt_offsets_diagram} is extremely flat for D16 (upper centre panel). 

The importance of these results will vary depending on science case. The overall small fraction of metallicities found so close to non-SF-flagged spaxels indicates that if the work being undertaken is statistical in nature, then these spaxels are unlikely to create substantial biases in the results. No matter the metallicity calibration in question, the sample of bordering metallicities is only about 1.3 per cent; for large samples, 1.3 per cent effects are unlikely to substantially impact results. 

Whether the goal is to remove the bias in the N2 based metallicities calibrations, or to remove the vast majority of these bordering spaxels for any metallicity sample, the imposition of the S06 diagnostic line is sufficient, leaving a sample of only around 1,350 spaxels out of a parent catalogue of $\sim$1.5 million spaxels. The 1,350 spaxels which remain also have a median offset of zero for all metallicity calibrations, which proved more effective in removing the bias than the inclusion of additional diagnostic diagrams.

\section{Conclusions}
\label{sec:conclusions}

In this work, we have undertaken an assessment of how frequently metallicities are found spatially adjacent to spaxels flagged as being dominated by radiation sources which are not consistent with photoionization, and undertaken a search for any systematic biases in metallicities which are found in these adjacent spaxels, and identified additional criteria to impose to remove any such biases.

\begin{itemize}
  \item We identify roughly 21,000 spaxels in approximately 1,200 galaxies with calculable metallicities which are found within 1 spaxel in $x$ and/or $y$ of a spaxel flagged as non-SF by the K03 BPT diagnostic line. We determine that such spaxels are relatively rare in the overall metallicity catalogue ($\sim$1.3 per cent), but are found in approximately 23 per cent of all galaxies with at least one metallicity. Exact values change slightly depending on the metallicity calibration used. 
  \item We use a suite of 6 metallicity calibrations, one each of the $R_{23}$, N2S2, and O3N2 based line ratios (KE08, D16, and PP04 O3N2 respectively) and a set of three \NII$_{6584}$/H$\alpha$ based calibrations (PP04 N2, M13 N2, and C17 N2) as the sample of examination in this work. 
  \item We construct a matched control sample of non-bordering spaxels for each identified border spaxel by identifying the 5 spaxels which are the closest match in radius, within a tolerance of $\pm$0.1 kpc in radius, with metallicities, and within the same galaxy. Our metallicity offsets are the difference between the metallicity of the border spaxel and the median of the matched controls. We also tested constructing a control sample matching in total stellar mass, surface mass density, and $R/R_e$, and find the results presented here qualitatively unchanged.
  \item We find that metallicities whose calibrations are based on $R_{23}$, N2S2, and O3N2 line ratios have offset distributions centred around zero, consistent with the control. While there is broad scatter around this zero offset peak (more than $\pm 0.4$ dex range), there is no systematic shift to higher or lower metallicities due to the adjacency with a non-star forming spaxel. This range in offsets is partially driven by atypical individual spaxels. 
  \item By contrast, metallicities using exclusively the \NII$_{6584}$/H$\alpha$ line ratio as the basis of the calibration show a systematic offset to higher metallicities than the control spaxels, with median offsets up to 0.041 dex higher than the controls. The \NII$_{6584}$/H$\alpha$ based calibrations show a range in offsets that is similar to the other three calibrations, and is driven by systematic offsets in a large number of galaxies.
  \item The inclusion of additional diagnostic diagrams, based on \OI~and \SII~lines, does not substantially change these results; more than 80 per cent of bordering spaxels are flagged as star forming by each of these additional diagnostics. Requiring all three diagrams (BPT + \SII + \OI) classify a given spaxel as SF does not remove the shift to higher offsets for  \NII$_{6584}$/H$\alpha$ based metallicities or reduce the scatter present among the offsets for any metallicity shown in the present work. The inclusion of the \OI~diagnostic diagram does result in a loss of sample size, as the \OI~line is not always detected with S/N $> 3$. 
  \item Imposing a stricter BPT cut, and shifting the requirement to S06 in lieu of K03 therefore reduces the sample size by $>$90 per cent down to approximately 1350 spaxels. The imposition of the S06 requirement removes the median offset to higher metallicities for all three of the \NII$_{6584}$/H$\alpha$ based metallicities, and does not impact the median offset of the $R_{23}$, N2S2, and O3N2 based metallicities, which remain at zero. This stricter cut does not substantially impact the range of possible offsets, which typically remains quite close to the nominal K03 selected sample. 
  \item We also find that imposing the \SII~and \OI~diagnostic diagrams does not offer any further reduction of either range or median offset once the S06 criterion has been implemented, though as before it does reduce the sample size. Requiring S06+\SII+\OI~diagnostics to all classify a spaxel as SF roughly halves the remaining border spaxels to approximately 350 spaxels, largely due to the inclusion of the \OI~diagnostic.    
    \item The overall scatter is not reduced as a result of these additional constraints, so individual non-SF-adjacent spaxels may still be offset by up to $\sim$0.25 dex. However, the small number of overall remaining non-SF-adjacent spaxels once the S06 BPT classification (with the optional inclusion of any others) is imposed means that these spaxels which remain are unlikely to affect any studies using statistical approaches within MaNGA. 
   \item Requiring the bordering spaxel to be doubly identified as SF (K03 BPT and \SII) and the non-SF spaxel to be doubly identified as non-SF reduces the scatter around the median metallicity offset by approximately half, but metallicities based on the N2 line ratio are still offset to positive values by about 0.02 dex.
   \item We also test whether the removal of the bias to higher metallicity values remains when the sample is selected based on S06 BPT classifications, rather than subsampling the K03-based sample. This sample is substantially larger than the K03 selected sample at $\sim66,500$ spaxels, but does not show any systematic bias in the metallicity values amongst the metallicity spaxels found adjacent to spaxels identified as non-SF by S06 on the BPT diagram. We conclude that the S06 diagnostic line functions well as an additional criterion to remove the bias seen in N2-based metallicity calibrations if they are adjacent to a spaxel flagged as non-SF in nature.
  \end{itemize}

We conclude that metallicities at the border of non-SF-flagged spaxels (identified using a BPT diagram cut similar to K03) are broadly consistent with their non-bordering spaxel counterparts within a galaxy, if the metallicities are based on $R_{23}$, N2S2, and O3N2 line ratios. \NII$_{6584}$/H$\alpha$ based metallicities are systematically shifted to higher metallicities by up to $\sim0.04$ dex. We interpret this sensitivity of the \NII$_{6584}$/H$\alpha$ based metallicities to adjacency of a spaxel flagged as non-star-forming as due to the bleed-through of a harder ionization field from a non-photoionization source, such as DIG or LI(N)ERs, which inflates the strength of the \NII$_{6584}$ line. With no other line ratios to rely on, the \NII$_{6584}$/H$\alpha$ are more sensitive to residual increases to partial ionization zones, even though the light captured in that spaxel may be dominated by photoionized emission. However, as the overall number of affected spaxels is low compared to the total number of metallicities possible to calculate within the MaNGA DR17, statistical studies are unlikely to be significantly affected by these biases. Case studies of individual galaxies, by contrast, may be more strongly affected. 

If science cases rely strongly on pure star forming regions, or if the results are sensitive to small number statistics, we therefore recommend the use of either a metallicity calibration that does not solely rely on the \NII$_{6584}$/H$\alpha$ line ratio, or to use a stricter BPT diagnostic cut to remove spaxels adjacent to non-SF-flagged spaxels from the sample.

\section*{Acknowledgements}
The authors thank the anonymous referee for their generous, constructive and careful review of this manuscript, which has improved the quality of this work.

JMS thanks Sara L. Ellison for constructive comments on a draft of this manuscript which improved the clarity of this work.

JMS acknowledges support from the National Science Foundation under Grant No 2205551. 
JMS also acknowledges the paid labour of undergraduate students not listed as coauthors, who worked on exploratory early phases of this work, or who worked with this data set to increase their skills with programming: Miles McClendon, Abby Tejera, and Megan Reilly. 

All plots were made using {\sc{matplotlib}} \citep{matplotlib}. We also made use of the {\sc{scipy}} \citep{scipy} module for statistical functions.

Funding for the Sloan Digital Sky Survey IV has been provided by the Alfred P. Sloan Foundation, the U.S. Department of Energy Office of Science, and the Participating Institutions. SDSS acknowledges support and resources from the Center for High-Performance Computing at the University of Utah. The SDSS web site is www.sdss4.org.

SDSS is managed by the Astrophysical Research Consortium for the Participating Institutions of the SDSS Collaboration including the Brazilian Participation Group, the Carnegie Institution for Science, Carnegie Mellon University, Center for Astrophysics | Harvard \& Smithsonian (CfA), the Chilean Participation Group, the French Participation Group, Instituto de Astrof\'isica de Canarias, The Johns Hopkins University, Kavli Institute for the Physics and Mathematics of the Universe (IPMU) / University of Tokyo, the Korean Participation Group, Lawrence Berkeley National Laboratory, Leibniz Institut f\"ur Astrophysik Potsdam (AIP), Max-Planck-Institut f\"ur Astronomie (MPIA Heidelberg), Max-Planck-Institut f\"ur Astrophysik (MPA Garching), Max-Planck-Institut f\"ur Extraterrestrische Physik (MPE), National Astronomical Observatories of China, New Mexico State University, New York University, University of Notre Dame, Observat\'orio Nacional / MCTI, The Ohio State University, Pennsylvania State University, Shanghai Astronomical Observatory, United Kingdom Participation Group, Universidad Nacional Aut\'onoma de M\'exico, University of Arizona, University of Colorado Boulder, University of Oxford, University of Portsmouth, University of Utah, University of Virginia, University of Washington, University of Wisconsin, Vanderbilt University, and Yale University.

%%%%%%%%%%%%%%%%%%%%%%%%%%%%%%%%%%%%%%%%%%%%%%%%%%
\section*{Data Availability}

The emission line data underlying this manuscript are publicly available as part of the MaNGA DR17 data release, available at \href{https://www.sdss.org/dr17/}{https://www.sdss.org/dr17/}. Metallicity values themselves are available upon reasonable request to the corresponding author.



\bibliographystyle{mnras}
\bibliography{Library.bib}% if your bibtex file is called example.bib

%\begin{thebibliography}{99}
%\end{thebibliography}

%%%%%%%%%%%%%%%%%%%%%%%%%%%%%%%%%%%%%%%%%%%%%%%%%%

%%%%%%%%%%%%%%%%% APPENDICES %%%%%%%%%%%%%%%%%%%%%

\appendix

\section{Additional figures}
\label{sec:appendix}

Here we present a few additional figures of the \SII~and \OI~diagnostic diagrams, and a plot indicating the correlation between metallicity offsets in non-SF adjacent spaxels and the emission line ratio \NII$_{6584}$/H$\alpha$. We also include a table indicating the exact breakdown of how the spaxels identified as non-SF by the K03 BPT diagnostic diagram, and which have spaxels with metallicities immediately adjacent to them, are classified on the BPT diagram, the \SII~diagnostic diagram, and the \OI~diagram. 

\begin{figure*}
	\includegraphics[width=0.82\textwidth]{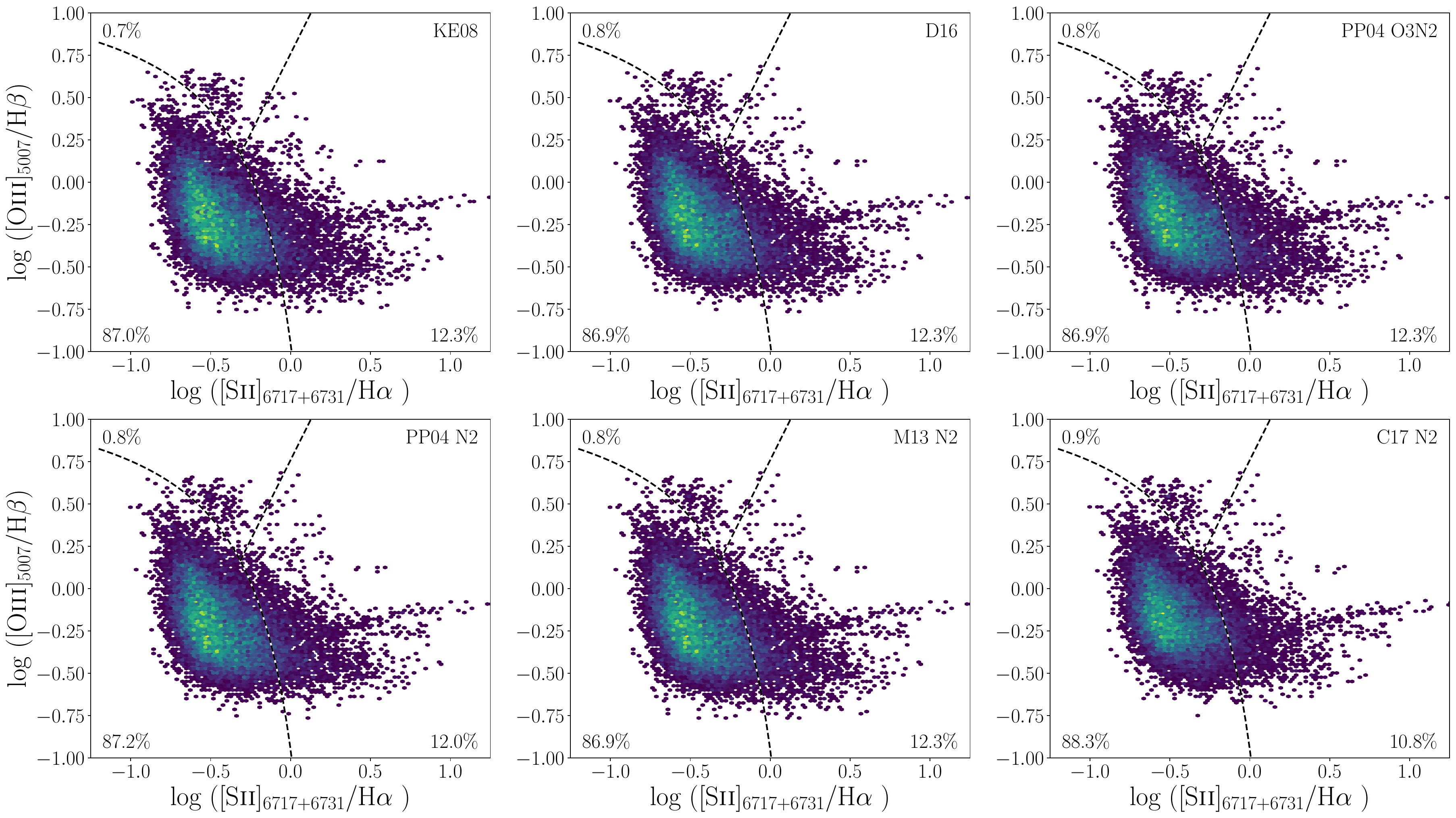}
    \caption{Log density histogram of the location of non-SF-bordering spaxels on \SII~based diagnostic diagram with division lines as presented in \citet{Kewley2006}.  The sample presented here is a subsample of the full sample, as only spaxels with \SII~detections with S/N $ > 5$ are plotted here, which reduces the sample of each metallicity by about 1,500 spaxels. We indicate the exact percent of the sample which falls into each section of the diagram in each panel. Typically, 0.8 per cent of the sample falls in the Seyfert regime, with 12 per cent in the LINER regime, and 87 per cent classified as SF.}
    \label{fig:SII_diagnostic}
\end{figure*}

\begin{figure*}
	\includegraphics[width=0.82\textwidth]{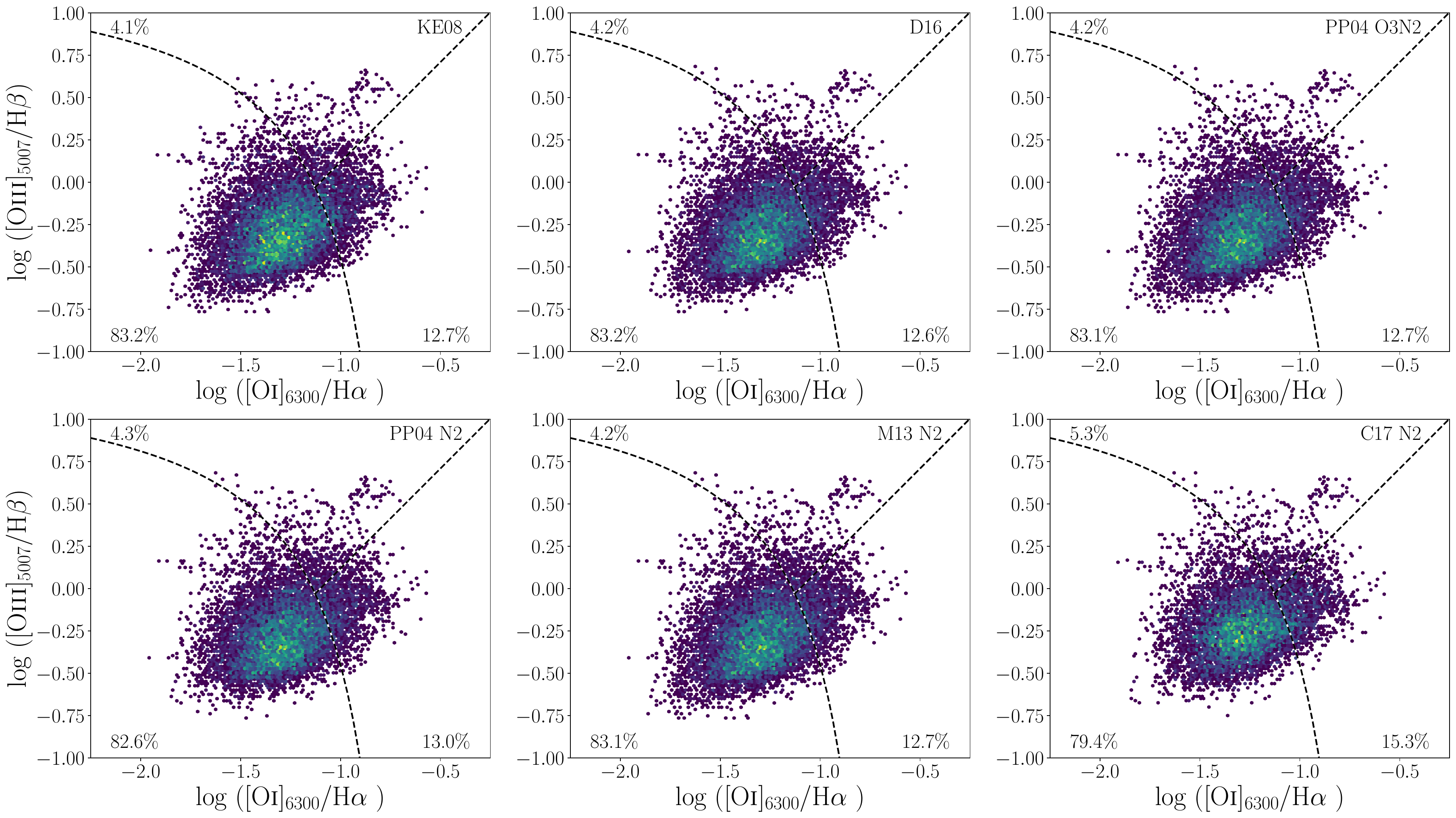}
    \caption{Log density histogram of the location of non-SF-bordering spaxels on \OI~based diagnostic diagram with division lines as presented in \citet{Kewley2006}. The subsample in this figure is slightly smaller than that of Figure \ref{fig:SII_diagnostic}, at 12,000 spaxels, or about 56 per cent of the parent sample due to the requirement of \OI~S/N $> 3.0$ for placement on the diagram. Around 83 per cent of border spaxels are classified as SF by this diagnostic, with 10 per cent flagged as a LINER, and about 5 per cent as a Seyfert AGN. }
    \label{fig:OI_diagnostic}
\end{figure*}

\begin{figure*}
	\includegraphics[width=0.82\textwidth]{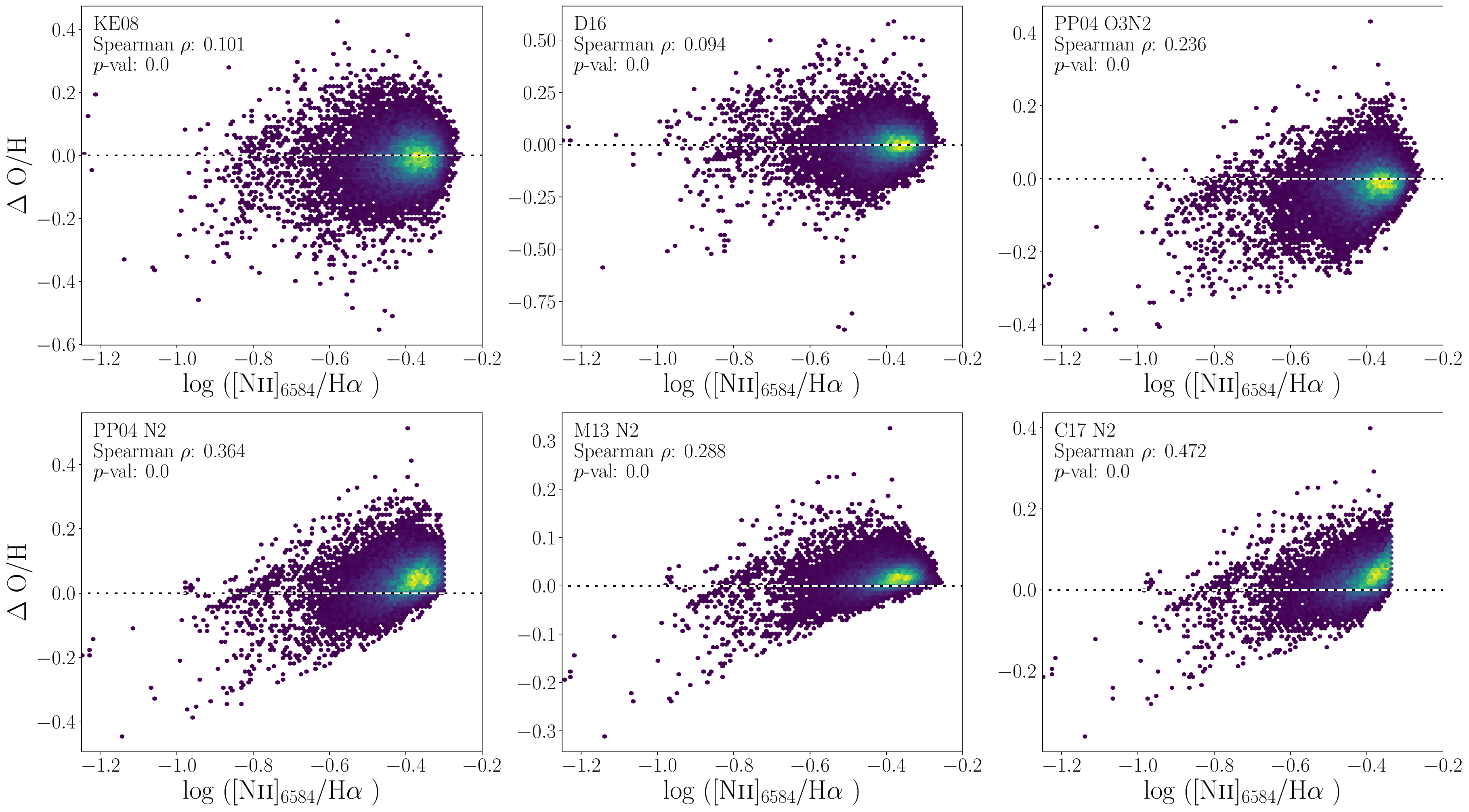}
    \caption{Log density histogram of the correlation between the \NII$_{6584}$/H$\alpha$ line ratio ($x$-axis) and the metallicity offset in the bordering spaxels, with one offset presented per panel, labeled in the bottom left corner.A horizontal black and white dashed line indicates the zero line of no offset. The top left corner has labeled the results of a Spearman Rank correlation test, where both the correlation coefficient and corresponding $p$-val are labeled. Spearman Rank correlation coefficients are quite low (less than 0.25) for all metallicities in the top row (KE08, D16, PP04 O3N2), while for the bottom panels, which are all metallicities based exclusively on the \NII$_{6584}$/H$\alpha$ line ratio, we find correlation coefficients of 0.288 (M13 N2), 0.363 (PP04 N2), and 0.471 (C17 N2), with both correlations and offsets to positive values visible by eye.}
    \label{fig:n2_elines}
\end{figure*}

\begin{table*}
	\centering
	\caption{For each metallicity calibration, we show the total number of spaxels with calculable metallicities, and the number of galaxies represented in that number of spaxels in the first and second columns respectively. In the third column we present the number of metallicity spaxels which are identified as bordering a non-SF spaxel as defined by the S06 BPT classification, and in the fourth, the percent of the total spaxel count that sample represents. In the fifth column we show the number of galaxies represented in our border spaxel sample, and in the 6th the percent of the number of metallicity hosting galaxies which have metallicities in our border spaxel sample. In the 7th and 8th columns, we show the mean and median number of border spaxels per galaxy.}
	\label{tab:s06_sample}
	\begin{tabular}{lcccc c c c c }
\hline 
Metallicity &  Total &  Total& Border & \% total & Galaxies  &  \% total & Mean  & Median  \\
calibration &  spaxels  &  galaxies & spaxels & spaxels &  w/border spaxels & galaxies & spax/galaxy &spax/galaxy\\\hline 
KE08 & 1,569,559 & 5,219 & 66,628 & 4.2\% & 2,087 & 40.0\% & 31.9 & 15 \\
D16 & 1,565,452 & 5,186 & 65,716 & 4.2\% & 2,067 & 39.9\% & 31.8 & 15 \\
PP04 O3N2 & 1,590,222 & 5,255 & 66,902 & 4.2\% & 2,087 & 39.7\% & 32.1 & 15\\
PP04 N2 & 1,586,127 & 5,253 & 66,947 & 4.2\% & 2,091 & 39.8\% & 32.0 & 15 \\
M13 N2 & 1,587,482 & 5,255 & 66,955 & 4.2\% & 2,091 & 39.8\% & 32.0 & 15\\
C17 N2 & 1,574,395 & 5,231 & 66,610 & 4.2\% & 2,077 & 39.7\% & 32.1 & 15\\
\hline 
\end{tabular}
\end{table*}

\begin{table*}
	\centering
	\caption{For each metallicity calibration, we show the total number of spaxels with calculable metallicities, and the number of galaxies represented in that number of spaxels in the first and second columns respectively. In the third column we present the number of metallicity spaxels which are identified as SF by both K03 and the \SII~diagnostic, and which are bordering a non-SF spaxel as identified as non-SF by both K03 and the \SII~diagnostic diagram. In the fourth, the percent of the total spaxel count that sample represents. In the fifth column we show the number of galaxies represented in our border spaxel sample, and in the 6th the percent of the number of metallicity hosting galaxies which have metallicities in our border spaxel sample. In the 7th and 8th columns, we show the mean and median number of border spaxels per galaxy.}
	\label{tab:doubleSF_sample}
	\begin{tabular}{lcccc c c c c }
\hline 
Metallicity &  Total &  Total& Border & \% total & Galaxies  &  \% total & Mean  & Median  \\
calibration &  spaxels  &  galaxies & spaxels & spaxels &  w/border spaxels & galaxies & spax/galaxy &spax/galaxy\\\hline 
KE08 & 1,569,559 & 5,219 & 1,386 & 0.1\% & 337 & 6.5\% & 4.1 & 3\\
D16 & 1,565,452 & 5,186 & 1,367 & 0.1\% & 330 & 6.4\% & 4.1 & 3 \\
PP04 O3N2 & 1,590,222 & 5,255 & 1,391 & 0.1\% & 337 & 6.4\% & 4.1 & 3\\
PP04 N2 & 1,586,127 & 5,253 & 1,351 & 0.1\% & 327 & 6.2\% & 4.1 & 3\\
M13 N2 & 1,587,482 & 5,255 & 1,391 & 0.1\% & 337 & 6.4\% & 4.1 & 3\\
C17 N2 & 1,574,395 & 5,231 & 1,083 & 0.1\% & 264 & 5.0\% & 4.1 & 3\\
\hline 
\end{tabular}
\end{table*}

\begin{table*}
	\centering
	\caption{In this table we present the number of spaxels identified as non-SF by the K03 BPT classification, and which are found to have adjacent metallicities. We show all possible combinations of classifications on the BPT, \SII, and \OI~diagnostic diagrams possible in this sample, for all six metallicity calibrations used in this work. In the bottom row of this table, we show the total number of spaxels with all three diagnostic classifications. The first three columns show the classification for each of the three diagnostic diagrams, and then in pairs of two, we indicate the number of spaxels and the percent of the total sample which fall into that combination of classifications. Each pair of two columns is for a single metallicity calibration, indicated at the top of the table. We note that more than half of the sample with all three classifications is flagged as K01 SF + SF + SF, and another quarter of the sample is contained within K01 + SF + LINER and K01 SF + LINER + SF (the second and third rows).}
	\label{tab:disambig}
\begin{tabular}{ccccccccccccccc}
\hline 
\multicolumn{3}{|c|}{Diagnostic} & \multicolumn{2}{|c|}{KE08} & \multicolumn{2}{|c|}{D16} & \multicolumn{2}{|c|}{PP04 O3N2} & \multicolumn{2}{|c|}{PP04 N2} & \multicolumn{2}{|c|}{M13 N2} & \multicolumn{2}{|c|}{C17 N2}\\ 
 BPT & [SII] & [OI] & Spaxels & \%  & Spaxels & \%  & Spaxels & \%  & Spaxels & \%  & Spaxels & \%  & Spaxels & \%  \\ 
\hline 
 K01 SF & SF & SF & 4,950 & 55.66 & 4,927 & 55.72 & 4,950 & 55.61 & 4,808 & 55.27 & 4,950 & 55.61 & 3,674 & 52.02\\ 
 K01 SF & LINER & SF & 1,489 & 16.74 & 1,473 & 16.66 & 1,489 & 16.73 & 1,431 & 16.45 & 1,489 & 16.73 & 1,051 & 14.88\\ 
 K01 SF & SF & LINER & 1,595 & 17.93 & 1,582 & 17.89 & 1,595 & 17.92 & 1,595 & 18.34 & 1,595 & 17.92 & 1,488 & 21.07\\ 
 K01 SF & LINER & LINER & 354 & 3.98 & 352 & 3.98 & 354 & 3.98 & 352 & 4.05 & 354 & 3.98 & 336 & 4.76\\ 
 K01 SF & AGN & SF & 7 & 0.08 & 7 & 0.08 & 7 & 0.08 & 7 & 0.08 & 7 & 0.08 & 7 & 0.1\\ 
 K01 SF & SF & AGN & 284 & 3.19 & 280 & 3.17 & 284 & 3.19 & 284 & 3.26 & 284 & 3.19 & 284 & 4.02\\ 
 K01 SF & AGN & LINER & 3 & 0.03 & 4 & 0.05 & 4 & 0.04 & 4 & 0.05 & 4 & 0.04 & 4 & 0.06\\ 
 K01 SF & LINER & AGN & 74 & 0.83 & 75 & 0.85 & 75 & 0.84 & 75 & 0.86 & 75 & 0.84 & 75 & 1.06\\ 
 K01 SF & AGN & AGN & 113 & 1.27 & 116 & 1.31 & 116 & 1.3 & 116 & 1.33 & 116 & 1.3 & 116 & 1.64\\ 
\hline 
 AGN & SF & SF  & 12 & 0.13 & 12 & 0.14 & 12 & 0.13 & 12 & 0.14 & 12 & 0.13 & 12 & 0.17\\ 
 AGN & LINER & SF & 0 & 0.0 & 0 & 0.0 & 0 & 0.0 & 0 & 0.0 & 0 & 0.0 & 0 & 0.0\\ 
 AGN & SF & LINER & 3 & 0.03 & 3 & 0.03 & 3 & 0.03 & 3 & 0.03 & 3 & 0.03 & 3 & 0.04\\ 
 AGN & LINER & LINER & 1 & 0.01 & 1 & 0.01 & 1 & 0.01 & 1 & 0.01 & 1 & 0.01 & 1 & 0.01\\ 
 AGN & AGN & SF & 0 & 0.0 & 0 & 0.0 & 0 & 0.0 & 0 & 0.0 & 0 & 0.0 & 0 & 0.0\\ 
 AGN & SF & AGN & 0 & 0.0 & 1 & 0.01 & 1 & 0.01 & 1 & 0.01 & 1 & 0.01 & 1 & 0.01\\ 
 AGN & AGN & LINER & 0 & 0.0 & 0 & 0.0 & 0 & 0.0 & 0 & 0.0 & 0 & 0.0 & 0 & 0.0\\ 
 AGN& LINER & AGN & 3 & 0.03 & 4 & 0.05 & 4 & 0.04 & 4 & 0.05 & 4 & 0.04 & 4 & 0.06\\ 
 AGN & AGN & AGN & 6 & 0.07 & 6 & 0.07 & 6 & 0.07 & 6 & 0.07 & 6 & 0.07 & 6 & 0.08\\ 
\hline 
\multicolumn{3}{|l|}{Total spaxels} & \multicolumn{2}{|c|}{8,894} & \multicolumn{2}{|c|}{8,843} & \multicolumn{2}{|c|}{8,901} & \multicolumn{2}{|c|}{8,699} & \multicolumn{2}{|c|}{8,901} & \multicolumn{2}{|c|}{7,062}\\ 
\hline 
\end{tabular}
\end{table*}

%%%%%%%%%%%%%%%%%%%%%%%%%%%%%%%%%%%%%%%%%%%%%%%%%%

% Don't change these lines
\bsp	% typesetting comment
\label{lastpage}
\end{document}